\newif \ifsigplan \sigplanfalse
\newif \ifdraft \draftfalse  %

\ifsigplan
\documentclass[cm]{sigplanconf}
\else
\documentclass{sig-alternate}
\fi
\usepackage{etex}

\usepackage{expdlist}

\newenvironment{itemize*}%
  {\begin{itemize}%
    \setlength{\itemsep}{3pt}%
    \setlength{\parskip}{0pt}}%
  {\end{itemize}}

\ifsigplan
\else
\usepackage[colorlinks=true,
            urlcolor=blue,
            citecolor=blue,
            linkcolor=blue,
            bookmarks=false,
            bookmarksnumbered,%
            backref=page,
            pagebackref=page,
            linktocpage=true,
            pdftitle={Win-Move is Coordination-Free (Sometimes), by: Daniel
            Zinn, Todd J. Green, Bertram Ludaescher},
            pdfauthor={Daniel Zinn, Todd J. Green, Bertram Ludaescher},%
            pdfsubject={Languages, Theory, Databases, Distributed Systems},%
            pdfkeywords={Datalog, distribution, relational transducer,
            monotonicity, semi-monotonicity},
            ]{hyperref}
\fi

\usepackage{amsfonts}
\usepackage{amssymb}
\usepackage{textcomp}
\usepackage{colortbl}
\usepackage{xspace}
\usepackage{alltt}
\usepackage{epigraph}
\usepackage{latexsym}
\usepackage{color}
\usepackage{subfigure}
\usepackage{wrapfig}
\usepackage{listings}
\usepackage{url}
\usepackage{tikz}
\usepackage[american]{babel}
\usepackage{algorithm}
\usepackage{fancybox}
\usepackage[normalem]{ulem}
\usepackage[hyperref,thmmarks]{ntheorem}  
\usepackage{verbatim} 

\usepackage{mytheorem}
\usepackage{local}

  \toappear{Permission to make digital or hard copies of all or part of
          this work for personal or classroom use is granted without fee provided that
          copies are not made or distributed for profit or commercial advantage and
          that copies bear this notice and the full citation on the first page. To copy
          otherwise, to republish, to post on servers or to redistribute to lists,
          requires prior specific permission and/or a fee. \par
          {\confname ICDT 2012}, March 26--30, 2012, Berlin, Germany.\par
          Copyright 2012 ACM 978-1-4503-0791-8/12/03 ...\$10.00}

\title{Win-Move is Coordination-Free (Sometimes)}

\ifsigplan
\conferenceinfo{ICDT '12}{}
\copyrightyear{2012}
\authorinfo{Daniel Zinn}
           {LogicBlox, Inc.}
           {dzinn@ucdavis.edu}
\authorinfo{Todd J. Green}
           {Department of Computer Science\\             
             University of California, Davis\\
             1 Shields Ave\\
             Davis, CA 95616}
           {green@cs.ucdavis.edu}
\authorinfo{Bertram Lud{\"a}scher}
           {Genome Center\\
             University of California, Davis\\
             1 Shields Ave\\
             Davis, CA 95616}
           {ludaesch@cs.ucdavis.edu}
\preprintfooter{\version}
\else
\numberofauthors{5}
\author{
\alignauthor
Daniel Zinn$^{1}$\\
       \email{daniel.zinn@logicblox.com}
\alignauthor
Todd J. Green$^{1,2}$\\
       \email{green@cs.ucdavis.edu}
\alignauthor Bertram Lud{\"a}scher$^{2}$\\
       \email{ludaesch@cs.ucdavis.edu}
\and
\alignauthor
       \affaddr{$^1$ LogicBlox, Inc.\\
         1349 W Peachtree St NW\\
        Atlanta, GA 30309  USA}
\alignauthor
       \affaddr{$^2$ Dept.~of Computer Science}\\
       \affaddr{University of California, Davis\\
       Davis, CA 95616  USA}\\
}
\fi

\ifsigplan
\newcommand{\version}{\lstinline|$Id: main.tex,v 1.255 2012/03/19 23:22:45 q Exp $|}
\else
\newcommand{\version}{\boxput{\makebox(-260,-30){{\textnormal{\lstinline|$Id: main.tex,v 1.255 2012/03/19 23:22:45 q Exp $|}}}{\relax}}}
\fi

\newcommand{\fa}{\ensuremath{\forall}}

\begin{document}
\maketitle 
\begin{abstract}

  In a recent paper by
  Hellerstein~\cite{Hellerstein10}, a tight relationship was
  conjectured between the number of strata of a \datalogneg\ program
  and the  number of ``coordination stages''  required for its distributed computation.
  Indeed, Ameloot et al.~\cite{Ameloot11} showed that 
  a query can be computed by a coordination-free relational transducer network iff it is \emph{monotone}, 
thus answering in the affirmative a variant of Hellerstein's CALM conjecture, 
based on a particular definition of coordination-free computation.  
In this paper, we present three additional models for declarative networking. In
these variants, relational transducers have limited access to the way data is
distributed. This variation allows transducer networks to compute more queries
in a coordination-free manner: e.g., a transducer can check whether a ground
atom $A$ over the input schema is in the ``scope'' of the local node, and then
send  either $A$ or $\neg A$ to other nodes. 

We show the surprising result that the query given by the well-founded semantics of the unstratifiable \emph{win-move} program is coordination-free in some of the models we consider. 
We also show that the original transducer network model~\cite{Ameloot11} and our
variants form a strict hierarchy of classes of co\-or\-di\-na\-tion-free
queries. 
Finally, we identify different syntactic fragments of \DLx, called \emph{semi-monotone} programs, which can be used as declarative network programming languages,
whose distributed computation is guaranteed to be eventually consistent and  co\-or\-di\-na\-tion-free.

\end{abstract}



\keywords{Datalog, distribution, relational transducer,
monotonicity\vspace*{-5mm}}

\section{Introduction}
\label{sec:introduction}

The popularization of cloud computing as a scalable, distributed computing paradigm has reinvigorated work on programming distributed systems and databases.  One of the thornier sources of inefficiency in such systems, identified first in~\cite{Hellerstein10} and reiterated in~\cite{Ameloot11,Koutris11}, is the %
presence of \emph{global coordination barriers} in distributed computation.

For example, consider a stratified \datalogneg program with a rule
of the form $A \leftarrow B, \neg C$. 
If a new fact $C$ becomes known, previously derived atoms $A$ may no longer be derivable, i.e.,  the program is \emph{non-monotonic}.
Therefore, the standard bottom-up evaluation
procedure for such programs involves a staged, stratum-at-a-time
evaluation~\cite{aliceBook}, with computation of a given stratum
allowed to proceed only when computation of all lower strata is complete.
In the distributed context, each stratum apparently constitutes a
synchronization barrier where all nodes must wait for all other
nodes from lower strata to finish.  Moreover, it seems that no results can be returned
to the user or client application until computation reaches the last
stratum.

In contrast, in Datalog programs without negation, the rules can be evaluated \emph{in any order} until a fixpoint is reached, since such positive programs are \emph{monotone}.
Even though this ``disorderly'' evaluation strategy can proceed in a  non-deterministic fashion, the resulting fixpoint is uniquely determined,  independent of the chosen evaluation order \cite{abiteboul1991Extension}. Thus, positive Datalog programs lend themselves to a co\-or\-di\-na\-tion-free evaluation. This has been exploited, e.g., by Loo et al.~\cite{Boon09} for declarative networking, using a pipelined semi-naive strategy, which, as Hellerstein~\cite{Hellerstein10} puts it, ``\emph{makes monotonic logic embarrassingly parallel}''. 
In contrast, stratified \datalogneg programs seem to require distributed coordination to proceed from one stratum to the next: ``\emph{Non-monotonic stratum boundaries are global barriers: in general, no node can proceed until all tasks in the lower stratum are guaranteed to be finished}'' \cite{Hellerstein10}.
Guided by this intution, Hellerstein conjectured:

\begin{Conjecture}[CALM~\cite{Hellerstein10}]
  A program has an eventually consistent, coordination-free execution
  strategy if and only if it is expressible in (monotonic) Datalog.
\end{Conjecture}

The CALM\footnote{Consistency And Logical Monotonicity} conjecture is informal---it requires precise notions of eventually
consistent, coordination-free, etc.---but seems plausible
nevertheless.  Ameloot et
al.~\cite{Ameloot11} have shown that  a version of the conjecture holds for transducer networks 
with a particular notion of coordination-freeness.
For stratified \datalogneg, 
Hellerstein~\cite{Hellerstein10}  suggests
that the parallel runtime ``\emph{might 
be best measured by the number of strata it must proceed through 
sequentially}'', which thus would yield a simple, elegant notion of coordination complexity, and further support the intuition that  ``\emph{there is deep connection between non-monotonic reasoning and parallel coordination}''.

A close correspondence between coordination complexity and number of strata in
\datalogneg programs would also nicely mimic the strict class hierarchy of
queries expressible by \datalogneg programs with $n$ strata: positive Datalog
programs are coordination-free and can be executed in any order within a single
stratum, while programs with $n$ strata could have a disorderly rule evaluation
within a stratum, but would require $n$ global coordination barriers between strata. 
Extrapolating these ideas even further, \emph{unstratifiable} \datalogneg programs, evaluated under the well-founded semantics~\cite{VanGelder91},  would then require an \emph{unbounded} number of coordination stages, corresponding to the number of rounds of the alternating fixpoint computation \cite{VanGelder93},
i.e., the number of coordination  barriers would no longer be fixed for a given program (as for stratified \datalogneg) but instead depend on the database instance.\footnote{Well-founded \datalogneg expresses the fixpoint queries~\cite{aliceBook}, and is strictly more expressive than stratified \datalogneg~\cite{Kolaitis1991}.}

In this paper, we revisit and shed new light on the CALM conjecture.
We show that, surprisingly, the canonical example of an unstratifiable program, the \emph{win-move} program,
has a coordination-free distributed evaluation strategy in some  of the declarative networking variants we consider. 

\subsubsection*{A Hierarchy of Relational Transducer Networks}

The original model, denoted \mB, are the transducer  networks defined by Ameloot et al.~\cite{Ameloot11}. Only monotone queries are coordination-free in this model.
In our first variant, \mP, transducers have some limited knowledge about the horizontal data distribution: 
a transducer can check whether an input atom $A$ (i.e., a ground fact $R(\bar
x)$ from the Herbrand base) is in its ``scope''. If that is the case, then the
local node can determine whether  $A$ or $\neg A$ is true using only local
computations, and pass on this information to other nodes. In \mP,  semi-positive \datalogneg programs are coordination-free.
In our second variant, \mR, we require partitioning policies to be \emph{element-determined}, i.e., there is a function $F$, mapping each domain element to a set of nodes. A ground input atom $R(x_1,\dots, x_k)$ is then in the scope of all nodes $F(x_i)$, for all $i=1,\dots,k$.
We show that in this model the unstratifiable win-move program  is coordination-free.

Finally, in  \mA, transducers have access to a system relation \adom containing the active domain of the global input.  As it turns out, this knowledge has a dramatic effect on the class of coordination-free programs, making all queries coordination-free.

Our main result, Theorem~\ref{thm:main}, states that the classes of queries that
are coordination-free under the network models \mB, \mP, \mR, and \mA, form a
strict hierarchy. 

\subsubsection*{{Games and Coordination}}

After presenting preliminaries in Section~\ref{sec:background}, we 
first motivate our approach for minimizing coordination in the presence of
non-monotonic constructs (negation), on the example of the well-known win-move
program (\secref{sec:winmove}).
We start from the ``doubled program'' \cite{KSS95} version of win-move and translate it into \DLx, which is a \datalogneg\ variant proposed by Abiteboul and Vianu~\cite{AbiteboulS91} with negation in the head (to indicate deletions) and $\forall$-quantification in the body.
The key idea of our approach to deal with the inherent difficulties in
programming distributed systems is to define a \emph{disorderly} semantics
for \DLx with built-in non-determinism that corresponds to non-determinism caused by
(i) network \emph{message re-ordering} or by
(ii) \emph{asynchronously} incorporationg incoming updates to the local
database state as in \cite{Boon09}. 
We show that our transformed program is confluent and terminating under this
semantics, and that the result agrees with both the deterministic and non-deterministic semantics of \DLx\ given by Abiteboul and Vianu.  Since the deterministic and non-deterministic semantics in general disagree for arbitrary  programs, another contribution is the identification of a ``well-behaved'' fragment of \DLx, which we call \datalogmon  (Section~\ref{sec:disorderly}).

Section~\ref{sec-transducer-networks} introduces the basic transducer network
model of Ameloot \etal \cite{Ameloot11}, called \mB here,  and then describes
our variants \mP, \mR, and \mA. 
Section~\ref{sec-coordination} presents our main result, the hierarchy of classes of coordination-free queries implied by these network models. The remaining sections contain discussions of related work and concluding remarks.

\section{Background and Preliminaries}
\label{sec:background}
\label{prelims}

\subsection{Datalog with Negation}

A \emph{relational schema} $\sigma$ consists of a finite
set of relation symbols $r_1,\dots,r_k$ with associated arities
$\alpha(r_i) \ge 0$.  Let \dom be a fixed and countable underlying
domain. A \emph{database instance (database)} over $\sigma$ is a
finite structure $$D=(U,r^D_1,\dots,r^D_k)$$ with finite universe
$U\subseteq \dom$ and relations $r^D_i \subseteq U^{\alpha(r_i)}$.
We will often identify database instances with sets of ground 
facts in the standard way, assuming $U$ to be the active domain.

A \emph{\datalogneg} program $P$ is a finite set of rules of the form
$$ A \leftarrow B_1,\dots,B_n,\neg C_1,\dots,\neg C_m$$
where the head $A$, positive body literals $B_i$ and negative body
literals $C_i$ are relational atoms, \ie of the form
$$R(x_1,\dots,x_l)$$ with $R \in \sigma_P$ being a relation symbol with
$\alpha(r) = l$ and each $x_i$ is either a variable or a constant from
\emph{dom}. The signature $\sigma_P$ of $P$ is partitioned into a set
\idb{P} of {\em intensional} relation symbols of $P$ occurring in some
head of $P$ and \edb{P} of {\em extensional} relation symbols
occurring only in the bodies of rules.  We require negation to be
safe, \ie in every rule, each variable must occur positively in the body. 
Among the relation symbols of
\idb{P}, one is distinguished as the {\em output predicate}.

A \emph{(positive) Datalog} program is a \datalogneg program that does
not have negative literals.  A \datalogneg
program $P$ is \emph{semi-positive} if all negatively used relations
are among \edb{P}.

Fix a \datalogneg program $P$. The \emph{predicate dependency graph}
for $P$ is the directed graph $\preddep(P) = (\sigma_P,E^+ \cup E^-)$
whose vertices are the relation symbols of $P$, and such that $(r,r')
\in E^+$ (resp.~$E^-$) if $r$ occurs positively (resp.~negatively) in
the body of a rule in $P$ that has $r'$ in the head.  $P$ is
\emph{recursive} if $\preddep(P)$ contains a cycle. $P$ is
\emph{stratifiable} if $\preddep(P)$ does not contain a cycle with a
negative edge $(r,r')\in E^-$. Every semi-positive \datalogneg program
is trivially stratifiable.

\TODO{conflict-resolution for deterministic semantics}

\TODO{Queries: distinguished output predicate}

\subsection{Well-Founded Semantics}
Let $P$ be a  \datalogneg\ program, including a given database \cD\ as
a set of facts.  Fix some Herbrand interpretation $\cJ \subseteq
\calB_P$, where $\calB_P$ denotes the set of all ground instances of
atomic formulas of $P$.  The {\em immediate consequences} under the
assumptions \cJ\ are given by:
\[
\begin{array}{l}
  \TP{\cJ}(\cI) := \\[1ex]
  \quad \{ H \ |\ (H \leftarrow B_1,\dots,B_n,\neg C_1,\dots,\neg C_m)
  \in \ground{P},\\[1ex]
  \quad\quad \cI \models B_1 \wedge \cdots \wedge B_n, \;\; \cJ \models \neg C_1
  \wedge \cdots \wedge \neg C_m \}.
\end{array}
\]
Since \cJ\ is fixed, $\TP{\cJ}$ is a monotone operator.  Let
$\Gamma_P(\cJ) := \lfp(\TP{\cJ})$ be its least fixpoint.  The operator
$\Gamma_P$ is antimonotone (observe how \cJ\ is used in $\TP{\cJ}$,
\ie $\cJ_1 \subseteq \cJ_2$ implies $\Gamma_P(\cJ_2) \subseteq
\Gamma_P(\cJ_1)$). It follows that $\Gamma^2_P (:= \Gamma_P \circ
\Gamma_P)$ is a monotone operator, so it has a least and a greatest
fixpoint.  These are used to define the 3-valued {\em well-founded
  model} $\cW_P$, a mapping of ground atoms to $\{ \true, \false,
\undef \}$ as follows:
$$ \cW_P(A) := \left \{ \begin{array}{ll} 
     \true & \textnormal{ if } A \in \lfp(\Gamma^2_{P,I}) \\
     \false & \textnormal{ if } A \not\in \gfp(\Gamma^2_{P,I}) \\
     \undef & \textnormal{ if } A \in \gfp(\Gamma^2_{P,I}) \setminus
     \lfp(\Gamma^2_{P,I}) 
     \end{array} \right.
$$

We recall also that the well-founded semantics is a conservative
extension of the stratified semantics, i.e., the well-founded
semantics is two-valued for stratifiable \datalogneg programs $P$; and
in these cases, the well-founded model agrees with stratified
model.

\subsection{Alternating Fixpoint} 
The construction of $\cW_P$ given above is called the {\em alternating
  fixpoint} computation of the well-founded model~\cite{VanGelder93} and
involves a nested fixpoint: The inner fixpoint is given by
$\Gamma_P(\cJ) = \lfp(\TP{\cJ})$, the outer fixpoints are obtained by
iterating the antimonotone operator $\Gamma_P$. The sequence
$\Gamma_P^0,\Gamma_P^1,$... given by
\begin{equation}\label{eq-gammai}
 \Gamma_P^0 := \emptyset \quad \textnormal{ and } \quad \Gamma_P^{i+1} :=
\Gamma_P(\Gamma_P^i), 
\end{equation}
alternates between underestimates of \true, and overestimates of \true
or \undef atoms, respectively. More precisely, the subsequence
$\Gamma_P^{2k}$ converges to the least fixpoint (the set of true
atoms) from below, while $\Gamma_P^{2k+1}$ converges to the greatest
fixpoint (the set of true or undefined atoms) from above. The key idea
is that applying the antimonotone operator $\Gamma_P$ to an
underestimate yields an overestimate and vice versa.

\subsection{Doubled Program}
A standard refinement of the alternating fixpoint
technique initializes the first underestimate with the
set of definitely true facts, \ie those that do not depend on any
negative subgoals~\cite{KSS95}. Furthermore, previously computed under- and
over-estimates are used to ``seed'' the fixpoint computation
exploiting monotonicity of $\TP{\cJ}(\cI)$ in $\cI$. Here, computing a
new underestimate starts from the old underestimate (since the
sequence of underestimates is monotonically increasing), and computing
a new overestimate starts from the last underestimate since
underestimates are always subsets of the overestimates. Let $P^+$
be the subset of rules of a program $P$ that do not contain
negation, and $\TPFP{\cJ}(\cI)$ denote the least fixpoint obtained
by iterating $\TP{\cJ}$ on $\cI$.  The {\em doubled program} approach
can be described as:
\[
\begin{array}{llll}
   \cU_0 & := & T_{P^+\!,\emptyset}^{\omega}(\emptyset) \\[0.75ex]
   \cV_0 & := & \TPFP{\cU_0}(\cU_0) \\[0.75ex]
   \cU_i & := & \TPFP{\cV_{i-1}}(\cU_{i-1}),  & i \ge 1 \\[0.75ex]
   \cV_i & := & \TPFP{\cU_{i}}(\cU_i), & i \ge 1
\end{array}
\]
This computation is performed until the sequence of underestimates
$\cU_k$ and overestimates $\cV_k$ become stationary. It is equivalent
with the definition given in 
(1) in the sense that
$\lfp(\Gamma^2_P)$ equals the fixpoint $\FP{\cU_i}{}$ and
$\gfp(\Gamma^2_P)$ equals the fixpoint $\FP{\cV_i}{}$.

We use the doubled program approach as a starting point for our
parallel win-move evaluation (\secref{sec:winmove}).  
However, note
that while in this scheme an underestimate $\cU_{i+1}$ is computed
``incrementally'' (using the previous underestimate $\cU_{i}$), an
overestimate $\cV_{i+1}$ is always computed ``almost from scratch'' (without
using the previous overestimate $\cV_i$ but only $\cU_i$), which is somewhat
inefficient.  
A number of 
papers (e.g., \cite{Bry89,DK89,ZukowskiF96,BrassDFZ01}) have explored
techniques for incrementally computing overestimates $\cV^{i+1}$ by directly inferring 
the ``to-be-deleted'' facts from $\cV^{i}$.

\subsection{Production Rules Semantics}  
A number of \emph{production rule} semantics have been defined for \datalogneg and
extensions
thereof~\cite{Vianu1997,AbiteboulS91,abiteboul1990non}.
These procedural semantics are easy to compute: The program is
understood as a set of rules that are fired until a fixpoint is
reached. A rule can fire if its body is true for the current database
instance. The rule heads are interpreted as \emph{updates}, \ie positive heads are insertions, while negative
literals are deletions.  Here, we consider
\DLx~\cite{AbiteboulS91}, a language whose syntax extends
\datalogneg\ by allowing negative literals in the head (deletions), and $\forall$-quantification of
variables in the rule body.  Thus, \DLx rules are of the form
\[
 A \leftarrow \forall \Xbar \ B_1, \dots, B_n, \neg C_1, \dots, \neg C_m
\]
where $A$ is a {\em positive} or {\em negated} ($\neg$) relational
atom\footnote{In the presentation of Abiteboul and
  Vianu~\cite{AbiteboulS91}, heads may have multiple atoms,
  giving additional expressive power under their non-deterministic
  semantics, but the more restricted version here is sufficient for
  our purposes.}, each variable not in \Xbar\ occurs in at least one
positive atom in the body, and the variables in \Xbar\ occur only in
the body and only in negated atoms.

It is convenient to allow \DLx\ programs to perform updates to
extensional relations; or, equivalently, to allow for some intensional
relations to come ``pre-initialized.''  
Thus we assume that the schema $\sigma_P$ of a
\DLx\ program $P$ is partitioned into a set \edb{P} of relation
symbols occurring only in the bodies of rules, and two sets \idb{P}
and \eidb{P} of relation symbols occurring in the heads of rules.  The
input to $P$ will be an instance consisting of \edb{P} and \eidb{P}
tuples.  The distinguished output predicate for $P$ must come from
 \idb{P}.

A \DLx\ program may be interpreted under either of two distinct
semantics, the deterministic and the non-deterministic semantics.
We will return to these semantics, and present a third ``even more
non-deterministic'' alternative which we call the disorderly semantics, in Section~\ref{sec:disorderly}.

Intuitively, the {\em deterministic} semantics involves firing all
applicable rules in parallel at each step of the computation.  Rules
with positive heads are treated as insertions, while those with
negative heads are considered deletions; conflicting insertions and
deletions are ignored.
The {\em non-deterministic} semantics for
\DLx, on the other hand, involves firing just a single ground rule at each
step of the computation, with the rule chosen non-deterministically.

Formally, given a \DLx\ program $P$ and database instance \cI, we
define the {\em deterministic immediate consequence operator} $\TPset$
as follows: 
\[
\begin{array}{ll}
  \TPset{\cI} := (\cI \cup ( \UpdateIns{\cI} \setminus \UpdateDel{\cI} ))
  \setminus (\UpdateDel{\cI} \setminus \UpdateIns{\cI} )
  \textnormal{ with}\\[0.75ex]
  \quad \Update{\cI} = \{ \, \nu(head) \; | \; \cI \models \nu(body) \textnormal{ for
  }(head \leftarrow body) \in P \,\}
\end{array}
\]
The result $\Pset(\cI)$ of applying a \DLx program $P$ to an
input instance $\cI$ under the \AVALL\ semantics is defined as the least fixpoint of iterating
$\TPset{\cdot}$ on $\cI$. In case the fixpoint does not exist
$\Pset(\cI)$ is undefined. Note, that for these procedural semantics we do
\emph{not} encode the EDB facts as body-less rules in $P$. Instead,
they are introduced by seeding the fixpoint computation. Since the
procedural semantics allow deletions, this is significant.

The non-deterministic semantics is based on the notion of {\em
  immediate successor}~\cite{Vianu1997} of a set of facts using a
rule, defined as follows.  Let $r = head \leftarrow body$ be a \DLx\
rule. Let $\cI$ be a set of facts and $\nu$ be a consistent variable
assignment for the variables in $r$ such that $\cI$ implies
$\nu(body)$. Then an instance $\cI'$ is an {\em immediate successor}
of $\cI$ if it can be obtained from $\cI$ by (a) deleting the fact $A$
if $\nu(head) = \neg A$, or (b) by inserting $A$ if $\nu(head) = A$.
An instance $\cJ$ is an {\em eventual successor} of $\cI$ using the
rules of $P$ if there exists a sequence $\cI_0 = \cI, \dots, \cI_n =
\cJ$ such that for each $i$, $\cI_{i+1}$ is an immediate successor of
$\cI_i$ using some rule in $P$.

\begin{Definition}
\label{def:NDresult}
Let $P$ be a \DLx\ program, and let \cI\ be a source instance.  The
{\em result} of applying $P$ to $\cI$ under the non-deterministic
semantics is the set containing all instances $\cJ$ such that $\cJ$ is an
eventual successor of $\cI$ using the rules of $P$, and there is no
immediate successor $\cJ' \neq \cJ$ of $\cJ$ using some rule in $P$.
\end{Definition}

\section{Basic Approach}
\label{sec:winmove}

In this section, we present the key intuitions for a parallel and
``disorderly'' evaluation strategy for the \emph{win-move game}:
\begin{equation} \label{eq:winmove}
  \fxbox{
  \lstinline|win(X) :- move(X,Y), !win(Y).| 
  }
\end{equation}
Here, we have a database instance over an active domain of {\em
  positions} and having a single binary {\tt move} relation.  A tuple
{\tt move(a,b)} can be read as indicating that ``from position {\tt a}
a player can move to position {\tt b}.''  In the game, two players $\White$
and $\Black$ take turns making moves, starting from a given position, with
$\White$ playing first.  A player {\em loses} at position $X$ if she cannot
move; and she {\em wins} at $X$ if she can move to a position which
the opponent loses.

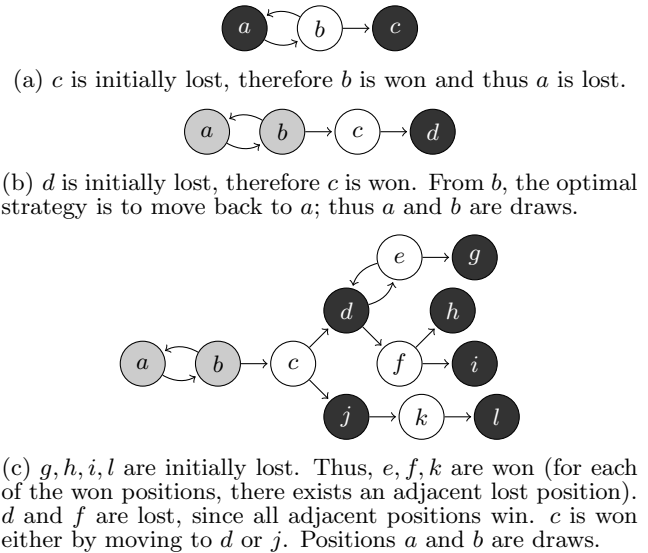
\begin{figure}[t]
\begin{center}
\subfigure[$c$ is initially lost, therefore $b$ is won and thus $a$ is lost.]{
  \parbox{0.977\columnwidth}{
  \centering
  \begin{tikzpicture}[shorten >=1pt,->]
    \tikzstyle{win}=[circle,draw=black,minimum size=17pt,inner sep=0pt]
    \tikzstyle{lost}=[circle,draw=black,fill=black!80,text=white,minimum size=17pt,inner sep=0pt]
    \tikzstyle{drawn}=[circle,draw=black,fill=black!20,minimum size=17pt,inner sep=0pt]
    \node[lost] (A) {$a$};
    \node[win, right of=A] (B) {$b$};
    \node[lost, right of=B] (C) {$c$};
    \path (A) edge[bend right] (B)
    (B) edge[bend right] (A)
    (B) edge (C);
  \end{tikzpicture}\\[1.5ex]
}
}
\subfigure[$d$ is initially lost, therefore $c$ is won. From $b$, the optimal
strategy is to move back to $a$; thus $a$ and $b$ are draws.]{
  \parbox{0.977\columnwidth}{
  \centering
  \begin{tikzpicture}[shorten >=1pt,->]
    \tikzstyle{win}=[circle,draw=black,minimum size=17pt,inner sep=0pt]
    \tikzstyle{lost}=[circle,draw=black,fill=black!80,text=white,minimum size=17pt,inner sep=0pt]
    \tikzstyle{drawn}=[circle,draw=black,fill=black!20,minimum size=17pt,inner sep=0pt]
    \node[drawn] (A) {$a$};
    \node[drawn, right of=A] (B) {$b$};
    \node[win, right of=B] (C) {$c$};
    \node[lost, right of=C] (D) {$d$};
    \path (A) edge[bend right] (B)
    (B) edge[bend right] (A)
    (B) edge (C)
    (C) edge (D);
  \end{tikzpicture}\\[1.5ex]
  }
  }
\subfigure[\label{winmovec} $g,h,i,l$ are initially lost. Thus, $e,f,k$ are
won (for each of the won positions, there exists an adjacent lost position). $d$
and $f$ are lost, since all adjacent positions win. $c$ is won either by moving to
$d$ or $j$. Positions $a$ and $b$ are draws. 
]{
  \parbox{0.977\columnwidth}{
  \centering
  \begin{tikzpicture}[shorten >=1pt,->]
    \tikzstyle{win}=[circle,draw=black,minimum size=17pt,inner sep=0pt]
    \tikzstyle{lost}=[circle,draw=black,fill=black!80,text=white,minimum size=17pt,inner sep=0pt]
    \tikzstyle{drawn}=[circle,draw=black,fill=black!20,minimum size=17pt,inner sep=0pt]
    \node[drawn] (A) {$a$};
    \node[drawn, right of=A] (B) {$b$};
    \node[win, right of=B] (C) {$c$};
    \node[lost, above right of=C] (D) {$d$};
    \node[lost, below right of=C] (J) {$j$};
    \node[win, above right of=D] (E) {$e$};
    \node[lost, right of=E] (G) {$g$};
    \node[win, below right of=D,yshift=0em] (F) {$f$};
    \node[win, right of=J] (K) {$k$};
    \node[lost, right of=K] (L) {$l$};
    \node[lost, above right of=F] (H) {$h$};
    \node[lost, right of=F] (I) {$i$};
    \path (A) edge[bend right] (B)
    (B) edge[bend right] (A)
    (B) edge (C)
    (C) edge (D)
    (C) edge (J)
    (D) edge[bend right] (E)
    (E) edge[bend right] (D)
    (E) edge (G)
    (D) edge (F)
    (F) edge (H)
    (F) edge (I)
    (J) edge (K)
    (K) edge (L);
  \end{tikzpicture}\\[1.5ex]
  }
}
\vspace{-6mm}
\end{center}
\caption{\label{winmoveexample}Examples of solved win-move games}
\vspace{-3mm}
\end{figure}

Evaluating program %
(2) under the well-founded
semantics, the \true\ facts in {\tt win} are the positions $X$ such
that \White\ has a winning strategy for the game starting at $X$,
while the \false\ facts in {\tt win} are the positions for which
\Black\ has a winning strategy.  The \undef\ facts in {\tt win} are
the {\em drawn} positions for which neither player has a winning
strategy, that is, the two players can move in cycles without either
one being able to force the other into a lost position.  See 
\figref{winmoveexample} for several examples of {\tt move} graphs
together with their solutions.

Our strategy uses a multi-step program transformation whose first step
is just the standard doubled program approach
(cf.~\secref{sec:background}).  Thus, we introduce two relations that
we shall call \wu\ and \wo, in which we compute the set of
\emph{definitely won} and \emph{not definitely lost} positions,
respectively. Here, \wu\ is the \emph{underestimate} of the {\tt win}
relation, while \wo\ is the \emph{overestimate} of the {\tt win}
relation.
\begin{center}
 $\,$\fbox{\lstinline{wu(X) :- move(X,Y), !wo(Y).}} $\;$ ($P_u$)\\[0.4ex]
 \fbox{\lstinline{wo(X) :- move(X,Y), !wu(Y).}} $\;$ ($P_v$)
\end{center}
\begin{Example}
  Consider \figref{winmovec}. The doubled program computation proceeds
  as follows to determine positions $c,e,f$ and $k$ as win; $a$ and
  $b$ as draws and the remaining positions $d,g,h,i,j$ and $l$ as lost
  positions.
\begin{center}
\ttfamily
\begin{tabular}{l|l}
 & \textnormal{estimates} \\
\hline
 $\cU_0$ & \wu\ = \set{} \\
 $\cV_0 $  & \wo\ = \set{a,b,c,d,e,f,j,k} \\
 $\cU_1 $ & \wu\ = \set{e,f,k} \\
 $\cV_1 $  & \wo\ = \set{a,b,c,e,f,k} \\
 $\cU_2 $ & \wu\ = \set{c,e,f,k} \\
 $\cV_2 $  & \wo\ = \set{a,b,c,e,f,k}
\end{tabular}
\vspace{1ex}
\end{center}
\end{Example}
Note that in the doubled program evaluation, $P_u$ and $P_v$ are
evaluated sequentially in a strictly alternating order. 
Directly mapping this approach to a distributed setting would seem to
require reaching a global consensus that a given step is complete
before the next step is allowed to proceed.  To avoid the need for
such coordination, we will derive a new set of rules in which all
rules can be applied to ground facts \emph{in any order}.  The new set
of rules will be a \DLx\ program, whose result turns out to agree with
the doubled program whether we evaluate it under the deterministic
semantics of Abiteboul and Vianu, under their non-deterministic
semantics, or under the disorderly semantics that we introduce in this
paper.

As a first step in this derivation, we introduce an auxiliary binary
relation \es\ in which we record moves that, so far as we know, may be
able to be played to avoid a certain defeat. Initially, \emph{any}
move \emph{out} of a node is such a move, and all interior nodes may
still win.  That is, we initialize our relations \wo\ and \es\ as
follows\footnote{The rule to compute \wo\ is equivalent to computing
  the first overestimate of the alternating fixpoint procedure.}:
\[
  \fbox{
    \boxup
      {good_move(X,Y) :- move(X,Y).}
      {may_win(X) :- move(X,Y).}
  } \quad (\wminit)
\] 
Now, intuitively, we can successively recognize won positions by
detecting an adjacent lost position, \ie one that may not win anymore
(or is not in the over-estimate).  Further, whenever a position has
been discovered to be won, moving into that position is definitely not
a good move since now the opponent would be left in a winning
position; thus the edge should be deleted from our \es\
relation. Finally, if no good moves remain from a position, then
it is certain that the position is lost.
\[ \;\;
\fbox{
\boxupppThree
  {*(1)space space won(X) :- move(X,Y), !may_win(Y).}
  {*(2)!good_move(X,Y) :- won(Y), move(X,Y).}
  {*(3)!may_win(X) :- forall Y !good_move(X,Y),move(X,_).}
}\;  (\wmdmon)
\]
\begin{Example}
  Continuing with the example of \figref{winmovec}, we now consider
  two possible computations according to the non-deterministic
  semantics for \DLx.  In either case, we repeatedly choose a
  valuation of variables that makes the body true for a rule in
  \wmdmon\ and the current database state, and apply the corresponding
  insertion or deletion.
  The two computations $A$ and $B$ are represented in
  \figref{fig:twoAlternateComputations} as follows: \m\ is an EDB
  relation, so it stays constant. The relation \es\ is initialized
  with the contents of \m, and during the computation we are
  subsequently deleting facts from \es. For \wo\ it is similar: \wo\
  is initialized with non-leaf nodes, some of which are going to be
  deleted during the computation.  \wu\ begins empty. The numbers in
  the columns $A$ (resp.~$B$) indicate the order in which a fact is
  derived into \wu\ or deleted from \es\ and \wo\ in the computation
  trace $A$ (resp.~$B$).  For example consider computation $A$.  The
  first three derivations performed are to successively declare
  positions \lstinline{e}, \lstinline{f}, and \lstinline{k} won
  according to rule (1) of \wmdmon.
  We can then apply rule (2) to delete \lstinline{(d,e)} from \es\
  since \lstinline{e} is lost (step 4), similar for \lstinline{(d,f)},
  and \lstinline{(j,k)}. Now, \lstinline{d} does not have a good-move
  out anymore, and thus can be deleted from \wo\ according to rule
  (3), same for \lstinline{j}. Finally, we can apply rule (1) again to
  declare \lstinline{c} won, since, for example, \lstinline{d} is not
  in \wo\ anymore.  Now, no applicable rule would change any of the
  IDB relations anymore.  We end up with the winning positions in
  \lstinline{won}; the drawn positions are those in \wo\ that are not
  also in \wu, in our case \lstinline{a} and \lstinline{b}.

In the computation $A$, we derive \lstinline{won(c)}
last\footnote{Observe how this order roughly corresponds to the
  alternating fixpoint order.}.  However, the different order $B$ of
evaluating \wmdmon\ derives \lstinline{won(c)} already after only four
rule applications. Intuitively, this is due to the fact that there is
a short ``winning path'' from $c$ through $j$, $k$, and $l$, which
does \emph{not} have to wait for 
the deletion of $d$ from \lstinline|may_win|.

Furthermore, both computations $A$ and $B$ reach the same result,
although \wmdmon\ contains negation and is non-monotonic.
\end{Example}

\begin{figure}
\begin{brelation}{move}{X}{Y}
 a & b \\
 b & a \\
 b & c \\
 c & d \\
 c & j \\
 d & e \\
 d & f \\
 e & d \\
 e & g \\
 f & h \\
 f & i \\
 j & k \\
 k & l \\
\end{brelation}
\hspace*{-0.5em}
\begin{blrelationTime}{\es}{X}{Y}
 a & b &  & \\
 b & a &  & \\
 b & c &  & \\
 c & d &  & \\
 c & j &  & \\
 \sout{d} & \sout{e} & 4 & 9 \\
 \sout{d} & \sout{f} & 5 & 7 \\
 e & d & & \\
 e & g &  & \\
 f & h &  & \\
 f & i &  & \\
 \sout{j} & \sout{k} & 6 & 2 \\
 k & l &  & \\
\end{blrelationTime}
\hspace*{-0.5em}
\begin{ulrelationTime}{\wo}{X}
  a  &  & \\
  b  &  & \\
  c  &  & \\
  \sout{d}  & 7 & 8 \\
  e  &  & \\
  f  &  & \\
  \sout{j}  & 8 & 3 \\
  k  &  & \\
\end{ulrelationTime}
\hspace*{-0.5em}
\begin{ulrelationTime}{\wu}{X}
e & 1 & 5 \\
f & 2 & 6 \\
k & 3 & 1 \\
c & 9 & 4 \\
\end{ulrelationTime}
\vspace{0.5ex}
\caption{\label{fig:twoAlternateComputations} Two non-deterministic evaluations
\lstinline{A} and \lstinline{B} of $\wmdmon$, solving the win-move
game in \figref{winmovec}.}
\end{figure}

The fact that computations $A$ and $B$ from the example evaluate
$\wmdmon$ differently, yet reach the same conclusion, is not a
coincidence: as we will show in Section~\ref{sec:disorderly}, it is a
consequence of the structure of $\wmdmon$ that holds for {\em any}
example of a win-move graph and order of evaluation.  Moreover, we
will also show there that the result is in accordance with the
well-founded semantics.

Since the order of picking rules for evaluation does not change the
final result, we have much more flexibility when we distribute the
computation, as discussed next.

\subsection{Distributed Execution}
We assume the contents of the relations \m, \wu, \wo, and \es\ to be
horizontally distributed across multiple nodes. A \emph{global}
database state thus only exists virtually as a union over all
\emph{local} databases.  We would like each node to be able to take
part in the computation, \ie execute rules and update the global
state. To do so, it is necessary that a node can judge by only
investigating its \emph{local} state whether the \emph{global}
database implies a certain valuation for a body of a rule. Consider a
rule that has only positive body literals, \eg \lstinline{a(X) :- b(X),c(X)}. Clearly, if a node $N$ has \lstinline{b(1)} and
\lstinline{c(1)} amongst its local data, than \lstinline{a(1)} can be
derived.  If \lstinline{a(1)} should be stored at a different node
$N'$, then an insertion request for \lstinline{a(1)} is sent to $N'$,
who in turn inserts it into its local database.  However, consider a
rule such as \lstinline{wu(X) :- move(X,Y),!wo(Y)}. To be applicable
with a binding of \lstinline{X}$=\!\!k$,\lstinline{Y}$=\!\!l$ on the
\emph{global} database state, it is in general not enough to check
whether the fact \lstinline{wo(}$l$\lstinline{)} is missing in the
local state, since it could be in one of the other partitions located
at a different node. Here, we exploit the fact that data is often not
partitioned arbitrarily, but for example, according to a
hash-function, or a range-partitioning scheme.
In our example here, we assume a
globally known, fixed partitioning function $h$ that maps each fact of the
active domain to a certain host. 
In our example, this simply means that each node is responsible for a certain
set of positions and its adjacent edges.
The tuple
\lstinline{wo(}$l$\lstinline{)}, when existing, is thus stored at node
$N := h($\lstinline{wo(}$l$\lstinline{)}$)$. Now, since node 
$N$ knows that \lstinline{wo(}$l$\lstinline{)} \emph{should} be stored
\emph{locally}, it can deduce that the valuation
\lstinline{X}$=\!\!k$,\lstinline{Y}$=\!\!l$ is applicable if
\lstinline{wo(}$l$\lstinline{)} is not present in its local database.

Each node applies valuations according to the program \wmdmon,
without coordination. Not only is the final result deterministic, but also can
the computation potentially proceed even if communication to other nodes has
been lost (or is slow). 
Returning to example \figref{winmovec}. Assume a partitioning function that
assigns positions $a,b,$ and $c$ to node $N_1$; $j,k,l$ to node $N_2$; and the
remaining positions to node $N_3$. 
Now, consider the case in which
the network is partitioned such that $N_3$ cannot communicate with the other
two nodes. Surprisingly, in this case, running program $\wmdmon$ on each node individually
with communication only between nodes $N_1$ and $N_2$ 
will compute the complete result! The intuitive reason is that the only
dependence between the subgraphs $a,b,c,j,k,l$ and $d,e,f,g,h,i$ is the edge
between $c$ and $d$. But then, the fact that
position $c$ is won, can be established from the fact that $j$ is lost and does
not require the cooperation of node $N_3$. From the view of nodes $N_1$ and
$N_2$, the final result of $d$ is irrelevant in deciding that $c$ is won.
Similarly, $d$ can be established to be lost by node $N_3$ independently.

\section{Disorderly Evaluation Model}
\label{sec:disorderly}

In this section, we formalize the ideas illustrated in the previous
section.  We proceed in three steps.  First, we define an ``even more
non-deterministic'' semantics for \DLx which we call the {\em
  disorderly} semantics.  The basic idea is to relax the
non-deterministic semantics by allowing deferred application of
updates, to model network delay effects.  Second, we identify a
syntactic fragment of \DLx, called \datalogmon, where the
deterministic, non-deter\-ministic, and disorderly semantics of \DLx\
coincide; and which is \emph{eventually consistent} in a certain
precise sense.  Third, we demonstrate that our construction for
evaluating win-move is eventually consistent and correct.

\subsection{The Disorderly Semantics}
\label{subsec:disorderlySemantics}

Intuitively, in a distributed system, each node only has local
knowledge about the global state.  As in declarative
networking~\cite{Boon09}, %
we assume that the ``global'' database is horizontally distributed accross various nodes
in the system.  Each node runs a copy of the \emph{same} \DLx\
program, but has access only to its own \emph{local} data.  During the
distributed computation, a node fires rules of the program based on
its locally available data, requests updates that are either applied
locally or shipped to other nodes.  Crucially, even rules using
negation in the body are fired using just locally available
information; correctness of the scheme will rely on source data being
partitioned in such a way that the local node always sees a
conservative underestimate of the relevant negative information,
making it safe to fire such rules.
To capture the network message delays and reordering which arise in
distributed systems, we include a global bag (multiset) of \emph{pending
  updates} in the description of the state of the distributed system,
manipulated by local nodes in a non-deterministic fashion.
Fix a \DLx\ program $P$ with schema $\sigma$.  An {\em update} is a
positive or negative ground literal over $eidb(P) \cup idb(P)$. 
A {\em state} of the computation is a pair $(\cI,
\cU)$ comprising a database instance $\cI$ over $\sigma$ and a bag
(multiset) $\cU$ of requested updates over $\sigma$.  \cU\ can be thought of as
a collection of pending or deferred updates.

Like the non-deterministic semantics, the disorderly semantics is based
on the notion of an {\em immediate successor}, but this time of states
rather than sets of facts.  Let $(\cI, \cU)$ be a state, let $r$ be a
\DLx\ rule,
\[
H \gets \forall \Xbar B_1, \dots, B_n,
\]
in $P$, and let $\nu$ be a consistent valuation of the free variables
in $r$ such that $\cI \models \forall \Xbar \nu(B_1) \wedge \cdots
\wedge \nu(B_n)$.  Then a state $(\cI', \cU')$ is an {\em immediate
  successor} of $(\cI, \cU)$ using $r$ if one of the following holds:
\begin{itemize*}
\item[] {\bf request} $\cI' = \cI$ and $\cU' = \cU \uplus \{ \nu(H)
  \}$, where $\uplus$ denotes bag union
\item[] {\bf insert} $A'$ is an update from $\cU$, $\cI' = \cI \cup
  \{ A'\} $, and $\cU' = \cU - \{ A' \}$, where $-$ denotes bag
  difference
\item[] {\bf delete} $\neg A'$ is an update from $\cU$, $\cI' = \cI
  \setminus \{ A' \}$, and $\cU' = \cU - \{ \neg A' \}$, where $-$ denotes bag
  difference
\end{itemize*}
A state $(\cJ, \cV)$ is an {\em eventual successor} of state $(\cI,
\cU)$ using the rules of $P$ if there exists a sequence $(\cI_0,
\cU_0) = (\cI, \cU), \dots, (\cI_n, \cU_n) = (\cJ, \cV)$ such that for
each $i$, $(\cI_{i+1}, \cU_{i+1})$ is an immediate successor of
$(\cI_i, \cU_i)$ using some rule in $P$.

\mbox
\\

Under the disorderly semantics, a program has a set of possible outcomes 
(due to non-deter\-min\-istic choices). Intuitively, the instance
$\cJ$ is part of this result set if a state $(\cJ,U)$ can be reached for which
the only further updates that can be requested or applied are ``no-ops.''

\begin{Definition}
\label{def:disorderly} 
Let $P$ be a \DLx\ program, and let $\cI$ be a source instance.  The
{\em result} of applying $P$ to $\cI$ under the disorderly semantics
is the set of all instances $\cJ$ such that there exists $\cV$
satisfying (i) $(\cJ, \cV)$ is an eventual successor of $(\cI,
\emptyset)$ using the rules of $P$, and (ii) $(\cJ, \cV)$ is a {\em
  terminal state}, i.e., it has no eventual successor $(\cJ', \cV')$
of $(\cJ, \cV)$ with $\cJ' \neq \cJ$.
\end{Definition}

The disorderly semantics is ``even more non-deterministic'' than the
non-deterministic semantics in the following sense:
\begin{Proposition} \label{prop:simulation} 
\ 

\begin{enumerate}
\item For any \DLx\ program $P$ and source instance \cI, if \cJ\ is in
  the result of $P$ under the non-deterministic semantics, then \cJ\
  is in the result of $P$ under the disorderly semantics.

\item There exists a \DLx\ program $P$, a source instance \cI, and an
  instance \cJ\ such that \cJ\ is in the result of $P$ under the
  disorderly semantics, but not under the non-deterministic semantics.
\end{enumerate}
\end{Proposition}
\begin{proof}
  (1) follows from the observation that any computation under the
  non-deterministic semantics can be emulated by strictly alternating
  update derivation and update application.  To prove (2), consider
  the \DLx\ program
\[
 P = \fbox{\boxupp
  {r :- !r, !s.}
  {s :- !r, !s.}
 } 
\]
applied to the empty source instance.  Under the non-deter\-ministic
semantics, the result is $\set{ \set{\mathtt{r}} , \set{\mathtt{s}}
}$, while under the disorderly semantics, the result is $\set{ \set{
    \mathtt{r} } , \set{\mathtt{s}}, \set{\mathtt{r},\mathtt{s}} }$.
\end{proof}

\subsubsection*{Practical Considerations}

For practical reasons, we are interested in \DLx\ programs $P$ that
compute a single, deterministic result, even while allowing the
computation of that result to be carried out in a non-deterministic or
disorderly fashion.  This is captured by the notion of a {\em
  functional fragment}~\cite{abiteboul1990non} of \DLx:

\begin{Definition} A \DLx program $P$ is \emph{functional}, under a
  given semantics, if the result under that semantics has cardinality
  $\leq 1$ for any source instance \cI.
\end{Definition}

However, even functional programs that compute exactly one result can
be undesirable from a practical point of view, when they admit
divergent evaluation sequences that never reach a terminal state.
Consider, for example, the following functional \DLx\ program:
\[
P = \; \fbox{\boxupp
   { space space t :- !a. -- p :- t.}
   {space space  a :- b. --  !p :- t.}
   }
\]
Evaluated under the disorderly semantics, the result of $P$ on \set{b}
is the single output \set{a,b}. Yet, if the first rule is used to
derive \lstinline{t} in a state $(I,U)$ (by for example, executing it
first), then no eventual successor of $(I,U)$ will satisfy requirement
(ii) of \defref{def:disorderly}.
This is problematic because we intuitively, would want to reach the
final result \emph{eventually} even though some ``bad choices'' have
been made while picking the order of evaluation.

Before defining a notion of termination, we first dispatch with one
technical issue.  Nothing in the formal semantics we have presented so
far rules out the possibility that, in a given evaluation trace, the
same update will be derived \emph{ad infinitum}, a given rule will
never be given a chance to fire, or a certain pending update will
never be applied.  To rule out such pathological cases, we first
introduce a notion of {\em fairness}, which informally requires that
each derivable update is eventually derived, and each derived update
is eventually applied.

\begin{Definition} \label{fairtrace}
  Given a \DLx program $P$. A sequence of states
  $(I^0,\emptyset),(I_1,U_1),\dots$ for which $(I^{i+1},U^{i+1})$ is
  an immediate successor using a rule in $P$ is a \emph{fair trace} if
  (1) for any state $(I^i,U^i)$ for which there is a rule $r = (head
  \leftarrow body) \in P$ and valuation $\nu$ with $I^i \models
  \nu(body)$, there are states $S^j = (I^j,U^j)$ and $S^{j+1}
  =(I^{j+1},U^{j+1})$ with $j \ge i$ such that $S^{j+1}$ is obtained
  from $S^j$ by requesting the update $\nu(head)$ (\ie $U^{j+1} = U^{j}
  \cup \set{\nu(head)})$.  And also, (2) if $A \in U^i$ ($\neg A \in
  U^i$), then there are states $S^j = (I^j,U^j)$ and $S^{j+1}
  =(I^{j+1},U^{j+1})$ with $j \ge i$ such that $S^{j+1}$ is obtained
  from $S^j$ by inserting (deleting) $A$.
\end{Definition}
Having ruled out these pathological traces, we can now define
termination\footnote{This property is also often called quiesence, \eg
\cite{Ameloot11,Hellerstein10}.}
desired:
\begin{Definition} 
  A program $P$ is \emph{terminating} under a given semantics if 
  every fair trace reaches a terminal state.
\end{Definition}
Finally, combining termination and determinism yields our desired
property of \DLx\ programs:
\begin{Definition} 
  A program $P$ is \emph{eventually consistent} under a given
  semantics, if it is functional and terminating under this semantics.
\end{Definition}

It was shown in~\cite{abiteboul1990non} that for \DLx\ under the
non-deterministic semantics, functionality is undecidable.  
  As might be expected, this property (along with
termination and eventual consistency) is undecidable for the
disorderly semantics as well:

\begin{Theorem}
  \label{thm:disorderlyTermination}
  Under the disorderly semantics, functionality, termination, and
  eventual consistency are undecidable for \DLx, even for programs
  without universal quantification or eidb relations.

\end{Theorem}

The proof is by reduction from the undecidable problem of checking
containment of (positive) Datalog programs~\cite{Shmueli93}, and can
be found in the Appendix.

\subsection{Semi-Monotone \DLx}

Consider again the win-move program from Section~\ref{sec:winmove},
which we saw there compiled into the \DLx\ program \wmdmon.  Observe
that the updates in the program follow a certain, regular form: in
particular, we only insert into the {\tt won} relation, while we only
delete from the {\tt good\_move} and {\tt may\_win} relations.
Moreover, {\tt won} occurs only positively in the bodies of rules,
while {\tt good\_move} and {\tt may\_win} occur only negatively.  We
are therefore motivated to define the following fragment of \DLx:

\begin{Definition} 
\label{def:dmon}
A \DLx\ program $P$ is \emph{semi-monotone} if the relation names in
\idb{P} occur only positively, while the relation names in \eidb{P}
occur only negatively.
\end{Definition}

Note that in the course of computation for such a program, the \IDBP\
relations only grow, while the \IDBM\ relations only shrink.  This
justifies the terminology \emph{semi-monotone \DLx} (or \emph{semi-monotone
Datalog} for short).
\begin{Theorem}\label{thm:functional}
  Every semi-monotone \DLx\ program is eventually consistent under the
  disorderly semantics.
\end{Theorem}
The proof can be found in the Appendix. 

We conclude the subsection by noting the following:

\begin{Corollary} 
\label{prop:concordance}
For semi-monotone \DLx\ programs, the deterministic,
non-deterministic, and disorderly semantics coincide.
\end{Corollary}

\begin{Proof} Let $P$ be a semi-monotone \DLx\ program, and let \cI\
  be a source instance.  Suppose that \cJ\ is the result of evaluating
  $P$ on \cI\ under the deterministic semantics.  By results of
  Abiteboul and Vianu~\cite{abiteboul1991Extension}, \cJ\ is in the
  result set of $P$ applied to \cI\ under the non-deterministic
  semantics.  By Proposition~\ref{prop:simulation}, \cJ\ is also in
  the result set of $P$ applied to \cI\ under the disorderly
  semantics.  But since $P$ is semi-monotone, it is functional under
  that semantics, by Theorem~\ref{thm:functional}.  It follows that
  the result under any of the three semantics is exactly \cJ.
\end{Proof}

\subsection{Correctness of the Transformed Win-Move%
}
\label{subsec:correctness}

Next, we return to the \DLx\ version \wmdmon\ of the win-move game
presented in Section~\ref{sec:winmove}.  Since \wmdmon\ is
semi-monotone, Theorem~\ref{thm:functional} tells us that it is
eventually consistent.  We now show that it also correctly computes
the result of the original \datalogneg\ under the well-founded
semantics:

\begin{Lemma} \label{winmovesemimonotone}
  Let $P$ be the \datalogneg\ version of the win-move game, let
  (\wminit,\wmdmon) be its semi-monotone \DLx\ translation, and let
  \cI\ be a source instance.  Denote by \cJ\ the result of applying
  first \wminit\ to \cI obtaining $\cI'$, and then \wmdmon\ to $\cI'$,
  both under the disorderly semantics.  Let $P'$ denote the extension
  of $P$ to include the facts of \cI.  Then for any ground fact {\tt
    win(a)} we have the following:
\[
W_{P'}(\win(a)) = \left\{
\begin{array}{lll}
\true &\text{ iff } &\wu(a) \in \cJ\\
\false &\text{ iff } &\wo(a) \not\in \cJ\\
\undef &\text{ iff } &\wu(a) \not\in \cJ \text{ and } \\
&& \wo(a) \in \cJ
\end{array}
\right.
\]
\NoEndMark
\end{Lemma}

\begin{Sketch} 
  The basic idea is to show that the alternating fixpoint computation
  is simulated by a certain disorderly computation.  (Since \wmdmon\
  is semi-monotone, it is functional by Theorem~\ref{thm:functional},
  hence any disorderly computation will produce the same result.)  The
  initialization stage computes the set of first overestimates.  The
  proof is then done by induction on the length of the alternating
  fixpoint computation $\Gamma^i$ using a strengthened induction
  hypothesis (which includes an added \es\ relation to the alternating
  fixpoint sequence).  The chosen execution for the disorderly
  semantics, computes all immediate consequences of the first rule in
  \wmdmon (after which the underestimates agree), then the second and
  the third (after which the overestimates agree).  Induction is done
  by proving the claim for $i=0,1$ as a base.  Proving for $i=2k+2$
  assuming $i=2k$ and $i=2k+1$, \ie correctness of a new underestimate
  is easy to show, since applying rule 1 to the earlier state
  naturally computes the new state. Correctness of the overestimates
  $i=2k+1$ is a little trickier and requires applying induction
  hypothesis for $i=2k,2k-1,2k-2$.  The key insight here is that the
  \es\ from both programs agree in the odd $i$, \ie in the
  over-estimations.
\end{Sketch}

\section{Transducer Networks}\label{sec-transducer-networks}

In this section we develop four closely-related models for distributed
computations.
The first model, denoted \mB, is that of relational transducer
networks as defined by Ameloot et al.~\cite{Ameloot11}.  The other
three models are new and vary from \mB along two dimensions: how the
input data is distributed across the network; and how much a
transducer node ``knows'' about the distribution process.  In the
model \mP, input facts are distributed according to a distribution
policy which assigns each fact of the Herbrand base over the
extensional schema to one or more nodes.  Further, the distribution
policy is known to each node in the transducer network, in a sense we
shall make precise shortly. In model \mR, the facts of a $k$-ary
relation are distributed with a replication factor of at most $k$, determined
by their attribute values. Finally, model \mA differs from \mP in that
each transducer also has information about the active domain of the
global input instance.

In the following, we describe these models in detail, beginning with a
review of transducers and networks of relational transducers (model
\mB).

\subsection{Background}

We follow the paper by Ameloot et al.~\cite{Ameloot11} in presenting
the background notions of relational
transducers~\cite{abiteboul1998relational} and relational transducer
networks~\cite{Ameloot11}.

\subsubsection*{Relational Transducers}

A \emph{transducer schema} is a tuple $(\Sin,\Ssys,\Smsg,\Smem,k)$ of
four disjoint database schemas along with an arity $k$.  (The
subscripts stand for `input', `system', `message', and `memory',
respectively.)  An \emph{abstract relational transducer}
(\emph{transducer} for short) over this schema is a collection of
queries $ \set{Q^R_\tnsnd \st R \in \Smsg} \cup \set{Q^R_\tnins \st R
  \in \Smem} \cup \set{Q^R_\tndel \st R \in \Smem} \cup
\set{Q_\tnout}$, where
\begin{itemize*}
\item every query is over the combined schema $\Sin \cup \Ssys \cup
  \Smsg \cup \Smem$;

\item the arity of each $Q^R_\tnsnd$, each $Q^R_\tnins$, and each
  $Q^R_\tndel$ equals the arity of $R$; and
  
\item the arity of $Q_\tnout$ equals the output arity $k$.
\end{itemize*}
Here, `snd' stands for `send'; `ins' stands for `insert'; `del' stands
for `delete'; and `out' stands for `output'. A {\em state} of the
transducer is an instance of the combined schema $\Sin \cup \Ssys \cup
\Smem$.  
A {\em message instance} is an instance of $\Smsg$. Such a message
instance can stand for a set of messages (facts) received by the
transducer, or a set of messages sent by the transducer; the intended
interpretation will always be clear in context.

Let $\mathcal T$ be a transducer. A {\em transition} of $\mathcal T$
is a five-tuple $(I,I_\tnrcv,J_\tnsnd,J_\tnout,J)$, also denoted 
$I,I_\tnrcv \stackrel{J_\tnout}{\longrightarrow}J,J_\tnsnd$, where $I$
and $J$ are states, $I_\tnrcv$ and $J_\tnsnd$ are message instances,
and $J_\tnout$ is a $k$-ary relation such that
\begin{itemize*}
  \item every query of $\mathcal T$ is defined on $I' = I \cup
    I_\tnrcv$; 
  \item $J$ agrees with $I$ on $\Sin$ and $\Ssys$;
  \item $J_\tnsnd(R)$, for each $R\in S_\tnmsg$, equals
    $Q^R_\tnsnd(I')$;
  \item $J_\tnout$ equals $Q_\tnout(I')$;
  \item $J(R)$, for each $R \in S_\tnmem$, equals $R'$ as follows. Let
    the set of insertions for $R$ be $R^+ := Q^R_\tnins(I') \setminus
    Q^R_\tndel(I')$, the set of deletions for $R$ be $R^- :=
    Q^R_\tndel(I') \setminus Q^R_\tnins(I')$.  Then, $R' := (R \cup
    R^+) \setminus R^-$.  (That is, conflicting updates are
    ignored.)  Note that this allows assignment $R' := Q$ to be
    implemented.
\end{itemize*}
The intuition behind the instance $I'$ is that $\mathcal T$ sees its
input, system and memory relations, plus its received messages.  The
transducer does not modify the input and system relations.  The
transducer computes new tuples that can be sent out as messages; this
is the instance $J_\tnsnd$. The transducer also outputs some tuples
(which cannot later be retracted); this is the relation $J_\tnout$.
Finally the transducer updates its memory by inserting and deleting
some tuples in its memory relations.

Transducers are parameterized by the language $\mathcal L$ in which
the queries are expressed. A UCQ-transducer, for example, uses unions
of conjunctive queries.  Note that transducer transitions are
deterministic, in constrast to those of transducer networks, discussed
next.

\subsubsection*{Transducer Networks}

Next we recall transducer networks (model \mB), in which relational
transducers are placed on nodes in a communicating network.  Here a
{\em network} is a connected (not necessarily complete), directed
graph of {\em nodes} $N \subsetneq \dom$.  By insisting that the graph is
connected, we ensure that information can (eventually) flow between
any two nodes.  A {\em transducer network} is a pair $(N,\mathcal T)$
where $N$ is a network and $\mathcal{T}$ is a relational transducer.

Operationally, each node in the network has a relational transducer
and a {\em receive buffer} of incoming messages.  In the initial
state, all memory relations and receive buffers of all nodes are
empty. The system relations contain useful information about the
transducer network and input distribution (we will formally describe
what this entails).  The input $I$ of schema $\Sin$ to the transducer
network is partitioned (possibly with replication) across the input
relations via a partitioning function $H$ that maps every node $n$ to
a subset of $I$, such that $I=\bigcup_{n\in N} H(n)$.

The (global) {\em state} of a transducer network is a function
$\mState$ mapping each node $n \in N$ to a pair $(I, B)$ where $I$ is
a (local) transducer state and $B$, the {\em receive buffer}, is a bag
of facts over the schema $\Smsg$.  The state of a transducer network
evolves via two kinds of transitions.  In a {\em delivery transition},
a node reads and removes one fact over the schema $\Smsg$ from its
input buffer $B$, adds the fact to the appropriate memory relation,
makes a local transducer transition transforming the local state $I$
to $J$, and sends the resulting message instance $J_\tnsnd$ to its
neighbors.  A \emph{heartbeat transition} is the same as a delivery
transition, but no input is read from the input buffer.  Sending
$J_\tnsnd$ to neighbors means that after the transition, $J_\tnsnd$ is
added to the input bags of the neighboring nodes.

\subsubsection*{Non-Determinism and Desired Properties}

To define desired properties of network transducers, we first require
the notion of a run.  A \emph{run} of a transducer network $(N,
\mathcal{T})$ on input $I$ according to a partitioning function $H$ is
an infinite sequence $(\tau_n)_n$ of transitions starting from an
input configuration with empty network buffers, empty memory relations, and input relations
populated according to the partitioning function $H$.  The
\emph{result} of a run is defined as the union over all $J_\tnout$
produced during the transitions of the run.  A run is \emph{fair} if
every node does heartbeat transitions infinitely often, and every fact
in every message buffer is eventually taken out by a delivery
transition.

Network transducer transitions are non-deterministic in several
respects: from a given state, many transitions are possible in general
depending on the choice of node where the transition occurs, the
choice of transition type (heartbeat or delivery), and, for delivery
transitions, the choice of fact delivered.  These correspond to the
kinds of non-determinism found in real distributed systems.

It is desirable nevertheless for transducer networks to produce the
same output regardless of the network topology, partitioning strategy,
or non-deterministic choices of the run.  We formalize this notion
below.

\begin{Definition} 
  A transducer $\mathcal T$ \emph{computes} the query $Q$ if for any
  input $I$, network $N$, and distribution $H$, the result of any fair
  run of $(N,\mathcal{T})$ on $I$ according to $H$ is $Q(I)$.
\end{Definition}

\subsection{Variations on Transducer Networks}
\label{modeldefs}

In basic relational transducer networks (model \mB), the horizontal
partitioning of the input data is done arbitrarily and, without
communication, nodes know only which part of the input data was
assigned to them.  By varying these assumptions, we derive 
three natural variations on the basic model.

In all of our variations, we derive the horizontal partition
function for a particular input $I$ from an instance-independent {\em
  partitioning policy}.  A {\em partitioning policy} for a schema
$\mathcal{SP}$ and network $N$ is a computable function \PartPol that
associates with each ground atom in the Herbrand base of $\mathcal{SP}$
a non-empty subset of the nodes of $N$.  The domain of the
function \PartPol is infinite (the policy is independent of a
particular input instance), covering all ``potential'' tuples.  Given
a partitioning policy \PartPol and an input instance $I$, we define
the horizontal partition function $H_{\PartPol,I}$ used for
distributing the input instance data as follows:
\[
  H_{\PartPol,I}(i) := \set{ f \st f \in I \textnormal{ and } i
    \in \PartPol(f) }  
\]
Note that for any horizontal partioning $H$ of an input instance $I$,
there is a partitioning policy \PartPol such that $H_{\PartPol,I} = H$.
(For one of the models to be described, however, we will restrict the
allowed partitioning policies such that this no longer holds.)

Next, we allow each transducer \emph{restricted access} to the
partitioning policy \PartPol by adding a relation \Local for each $R
\in \Sin$ to its system relations. \Local has the same arity as $R$,
and a tuple $\bar x$ is in \Localn (the copy of \Local at node $n$)
iff $n \in \PartPol(R(\bar x))$.  Intuitively, on a node $n$, there is
a tuple $\bar x \in$ \Local if $n$ is ``responsible'' for this tuple.
If $R(\bar x)$ is in the global input $I$, then $R(\bar x)$ will be
distributed to node $n$ (and possibly others).  Conversely, if node
$n$ finds $R(\bar x)$ absent from its local input, then $n$ ``knows''
that $R(\bar x)$ is {\em not} in the global input $I$.

Note that the \Locali are in general infinite relations; in practice
each node would be equipped with a decision procedure to check whether
an arbitrary tuple is contained in \Locali or not.  Also, while the
queries of a transducer at node $i$ may access \Locali, we require
that the queries still produce finite results.  (For UCQ-transducers,
for instance, this can be ensured by extending the notion of safety to
require all variables occuring in \Local atoms to also occur
positively in normal atoms.)  This requirement could be lifted by
restricting \Locali to the active domain of the global input instance,
but we prefer not to do this as we shall see in \lemmaref{thm:adom}
that providing nodes knowledge of the active domain has a dramatic
impact on the notion of coordination-freeness.

We are now ready to define our first variation on relational
transducer networks.
\begin{Definition} 
  An {\em \mP-$\mathcal L$-transducer network} is an
  $\mathcal{L}$-trans\-ducer network along with a partitioning
  policy \PartPol, in which each transducer is additionally provided
  system relations \Local for all $R \in \Sin$ as described above.
\end{Definition}

In our second model, we restrict the allowed partitioning policies to
those which, intuitively, map domain elements (rather than ground
facts) to nodes.  This captures the style of distribution used in our
construction for the win-move game in the earlier sections.  More
precisely, a partitioning policy $\PartPol$ is called {\em
  element-determined} if there exists a (unique) mapping $F : \domain
\rightarrow 2^N$ of domain elements to sets of nodes such that
\[
\PartPol(\, p(x_1,\dots,x_n)\,) = \bigcup_{i=1,\dots,n} F(x_i).
\]
for every ground fact $p(x_1,\dots,x_n)$ in the Herbrand
base of the transducer's input schema $\Sin$.

\begin{Definition} \label{defmptransducer}
  An {\em \mR-$\mathcal L$-transducer network} is an \mP-$\mathcal
  L$-trans\-ducer network whose associated partitioning policy is
  element-determined.
\end{Definition}
We do not impose any restriction on $\PartPol$ for nullary relations beyond
what was done for type $\mR$ transducer networks.
Note that the nodes of an \mR-$\mathcal
L$-transducer network have, essentially, full knowledge of the
underlying mapping $F : \domain \rightarrow 2^N$ for the partitioning
policy via their \Local relations, since $F(a) = $ \Local$\!\!(a, \dots,
a)$ for any domain value $a$ and (non-nullary) relation $R$.

Finally, our third variation on transducer networks exposes global
knowledge of the active domain to nodes.

\begin{Definition}
  An {\em \mA-$\mathcal{L}$-transducer network} is an
  \mP-$\mathcal{L}$-transducer network in which the transducers
  additionally have access to a system relation \adom containing the
  active domain of the global input instance.
\end{Definition}

For any of these models, we also require that a transducer is
``oblivious'' to the network and partitioning policy, in the sense
that the same transducer should produce the correct result on any fair
run, regardless of the choice of network and partitioning policy.
\begin{Definition} 
  A $\mathcal L$-transducer $\mathcal T$ \emph{computes} a query $Q$
  in model $X \in \{\mP, \mR, \mA\}$ if for every network $N$, every
  distribution policy \PartPol compatible with $X$, and every input
  instance $I$, every run of $\mathcal T$ in which the input is
  distributed according to $H_{\PartPol,I}$ results in the output
  $Q(I)$.  In this case we say that $\mathcal{T}$ is {\em consistent}.
\end{Definition}

\subsection{Disorderly Semantics via Transducer Networks \label{rundisorderlyontransducers}
}

In this section we will show how certain syntactic classes of
semi-monotone programs can be ``compiled down'' to equivalent
transducer networks of various kinds.  The semi-monotone program we
have given for computing the win-move game will be seen to obey the
syntactic restrictions of one of these classes.

\TJ{why not just use a semimonotone-transducer network?}

\TJ{need to be careful to distinguish between transducers and
  transdcuer networks in below}

\begin{Lemma} \label{thmwoforall}
  Let $Q$ be a query computed under the disorderly semantics by a
  semi-monotone, \fa-free \DLx program $P$ in which
  each rule has at most one negated eidb atom.  Then there exists a
  type $\mP$ $\UCQneg$-transducer network which computes $Q$.
\end{Lemma}

A similar result holds for type \mR transducer networks. 
Here, we are allowed to use multiple negated eidb in the body of rules and 
\fa-quantification. 
  A \DLx rule $r$ is \emph{friendly} if any negated atoms in the
body of $r$ share a common variable $x$ which is not
universally-quantified.

\begin{Lemma}
  \label{thmwforall}
  Let $Q$ be a query computed by a semi-mon\-o\-tone \DLx program $P$
  under the disorderly semantics.  If all rules in $P$ are friendly, then there
  exists a type $\mR$ FO-transducer network which computes $Q$.
\end{Lemma}

Even certain kinds of pre-processing are allowed: A Datalog program
consisting only of constant-free projection rules, \ie of the form
\lstinline{R'(barZ) :- R(barX)} with $\emptyset \neq \bar Z \subseteq
\bar X$, and all $X_i \in \bar X$ being variables,
is called a \emph{projection program}.
\begin{Lemma} \label{thmcomposition}
  Let $P$ be a projection program, and let $Q$ be a query as defined
  in Lemma~\ref{thmwforall}.  Then there exists a type $\mR$
  FO-transducer network which computes $Q \circ P$. %
\end{Lemma}
As is customary, we mean query compostion with $\circ$, \ie for any input
instance $I$, $Q \circ P := Q(P(I))$.
We point out that the programs $\wminit$ and $\wmdmon$ for computing win-move presented in
Section~\ref{sec:winmove} satisfy these requirements.
\\

\noindent \textbf{Proofs.} 
All proofs for these lemmata are constructive.
The intuition is that the created transducer networks
essentially perform a distributed computation of the disorderly semantics.
Each transition of the transducer implements the FO-query denoted by the body of
the semi-monotone program applied to its \emph{local state}. 
Derived updates are locally applied as well as 
broadcasted to all nodes in the network.
Eventual consistency for the disorderly semantics of semi-monotone programs guarantees 
the consistency of the transducer. 
The syntactic restrictions and access to the \lstinline{Local} relations 
are necessary to allow the evaluation of rule bodies in a 
distributed manner: Even though each transducer only has partial
\emph{local knowledge} of the virtual \emph{global database state}, the restrictions guarantee that 
(1) each conclusion drawn from the local state could also have been drawn from the
global state, and (2) each conclusion that can be drawn from the global state
can also be drawn on at least one node with only its local knowledge. 

The transducer $\mathcal T$ constructed for \lemmaref{thmcomposition} is a modification of
the one constructed for \lemmaref{thmwforall} that first computes the
projection-program locally. Since the partitioning policy is element-determined,
all necessary \lstinline{Local} system relations for the $idb(P)$, which are the
edb and eidb for the transducer, can be emulated. 

While the
proofs provide valuable insights in how to program the transducer to  
guarantee consistency and correctness, the
details are somewhat technical. We thus decide to skip them here and refer the
interested reader to Appendix~\ref{appendixproofsimulation}.

\section{Coordination}\label{sec-coordination}

Again following Ameloot et al.~\cite{Ameloot11}, we say that an
$\mathcal{L}$-transducer $\mathcal T$ that computes $Q$ is
\emph{coordination-free with respect to model \mB} if for every input
$I$ and every network $N$, there exists a distribution $H$ for which
when the transducer network $(N, \mathcal T)$ is run with only
heartbeat transitions, it already produces the correct result.  Note
that since local heartbeat transitions are deterministic, the result
is deterministic for a given input $I$, network $N$, and partitioning
$H$. A query $Q$ is \emph{coordination-free in \mB} if there exists a
coordination-free $\mathcal{L}$-transducer $\mathcal T$ that computes
$Q$, for some query language $\mathcal{L}$.

A main result of the paper by Ameloot et al.~\cite{Ameloot11} relates
coordination-freeness and monotonicity:
\begin{Theorem}\textnormal{\textbf{\cite{Ameloot11}}}
\label{thm:ameloot}
A query is coordination-free in model \mB if and only if it is
monotone.
\end{Theorem}

The notion of coordination-free queries extends naturally to models
\mP, \mR, and \mA (replace horizontal distributions with distribution
policies in the definitions above for \mB).  For $X \in
\{\mB,\mP,\mR,\mA\}$, denote by $\mathcal{F}[X]$ the class of
queries that are coordination-free with respect to model $X$.  Denote by
$\mathcal{C}$ the class of all computable queries, by $\mathcal M$
the class of monotone queries, and by $\mathcal{SP}$ the class of all queries
computable by a semi-positive \datalogneg program. \TODO{is this defined in perlims?}

The main result of this section is that these classes form the
following hierarchy:
\begin{Theorem}
\label{thm:main}
\[ \mathcal M = \mathcal F[\mB] \subsetneq
   \mathcal F[\mP] \subsetneq
   \mathcal F[\mR] \subsetneq
   \mathcal F[\mA] = \mathcal{C}
\]
\end{Theorem}
The first equality is just a restatement of Theorem~\ref{thm:ameloot};
and it is clear from the 
construction of the models that
$\mathcal{F}[\mB] \subseteq F[\mP]$, $F[\mP] \subseteq F[\mR]$, and
$F[\mP] \subseteq F[\mA]$.  The remainder of the section is devoted to
a proof that these inclusions are proper, and indeed that the above
hierarchy holds.

\subsection{Proof of Main Theorem}

First, we show that endowing the nodes with knowledge of the global
active domain has a dramatic impact on the class of coordination-free
queries:

\begin{Lemma} 
\label{thm:adom}
  Every computable query is coordination-free in model \mA, i.e.,
  $\mathcal{F}[\mA] = \mathcal{C}$.
\end{Lemma} 
This implies in particular that $\mathcal{F}[\mR] \subseteq
\mathcal{F}[\mA]$.
\begin{Proof}
  First, we claim that there is a coordination-free
  Datalog$^{\neg}$-transducer $\mathcal T$ of type \mA such that, on any network, and
  any input $I$ distributed according to any policy, any fair run
  reaches a configuration where every node has a local copy of the
  entire instance $I$ in its memory, and an additional flag
  \lstinline{ready} (implemented by a nullary memory relation) is
  true.  Moreover, the flag \lstinline{ready} does not become true at a node
  before that node has the entire instance in its memory.

  We use the programming technique \emph{broadcast} as described in 
  Appendix~\ref{appendixproofsimulation} 
  to let transducers send messages (containing update requests)
  that will eventually 
  be delivered to any node in the system; besides sending the updates out,
  they are also locally applied.
  The transducer does the following on heartbeat transitions: for each
  edb-relation \lstinline{R}, it uses \lstinline{adom} and \Local to
  broadcast existence and non-existence of tuples which is then stored
  in the memory relations \lstinline{R___m}, and
  \lstinline{R_m^-1}, respectively. With \lstinline|@all| denoting broadcast:
  \[
  \boxupp
   { R___m@all(barX) :- R(barX).} 
   { R_m^-1@all(barX) :- adom(X_1),...,adom(X_n),Local_R(barX),!R(barX).}
  \]
  Then, each edb relation \lstinline{R} also has a flag \lstinline{readyR}
  that is set as follows:
\begin{lstlisting}
 knownR(barX) :- R___m(barX).
 knownR(barX) :- R_m^-1(barX).
 not___readyR :- adom(X_1),...,adom(X_n),!knownR(barX).
 readyR :- !not___readyR. 
 ready :- readyR, readyS,... 
\end{lstlisting}
\TODO{wtf do I get latex errors when I add the last line to the Datalog programm???}
  Finally, \lstinline{ready} is implemented as a conjunction of all
  \lstinline{readyR} for each edb relation \lstinline{R}.
  This \datalogneg program is run in each transition by $\mathcal T$.

  Next, for any query $Q$ expressible in a query language $\mathcal L$,
  let $\mathcal L'$ be a query language at least as expressible as both
  $\mathcal L$ and stratified \datalogneg. The $\mathcal L'$-transducer 
  $\mathcal T$, which distributedly 
  computes $Q$, does now first use the sub-routine from above to distribute $I$ to all
  nodes, and then, on each node, when \lstinline{ready} is true, computes $Q$ in
  one step.
  
  $\mathcal T$ is coordination-free. For any network, consider the policy that
  allocates all data to a single node $n_0$. Even though none of the sent messages arrive, 
  $n_0$ will reach the ``\lstinline{ready}-state'' and  
  output the complete result.
\end{Proof}

\begin{Lemma}
  In model \mP, for each semi-positive \datalogneg program there exists an
  equivalent coordination-free \UCQneg-transducer.  As a consequence,
  $ \mathcal{SP} \subseteq \mathcal{F}[\mP]$, and in particular $\mathcal{F}[\mB]
  \subsetneq \mathcal{F}[\mP]$.
\end{Lemma}
\begin{Proof}  
Each semi-positive \datalogneg program $P$ is also a semi-monotone
program, which additionally satisfies the syntactic restrictions identified in
\lemmaref{thmwforall}. Let $\mathcal T$ be the type \mP transducer 
as in the
proof for \lemmaref{thmwforall}. We observe that $\mathcal T$ is
coordination-free. Consider again the partitioning policy mapping all data to
a single node $n_0$. $\mathcal T$ on $1$ will run the complete semi-monotone 
program already with only heartbeat transitions.
\end{Proof}

\begin{Lemma}
\label{lem:winmoveIsNTwo}
The query computing the won positions of the win-move game is \mR-coordination
free.
\end{Lemma}
\begin{Proof}
\lemmaref{winmovesemimonotone} states that $\wmdmon \circ \wminit$ computes the
won positions for win-move. \wmdmon\ satisfies the restrictions of
\lemmaref{thmwforall}; and $\wmdmon$ is a projection program. Now, consider the
\mR-FO-transducer $\mathcal T$ that was constructed in the proof for
\lemmaref{thmcomposition}. $\mathcal T$ is coordination-free: consider the
partitioning policy that puts all data on a single node $n_0$.
\end{Proof}

\begin{Lemma}
  \label{lem:winmoveNotNOne}
$\mathcal F[\mP]$ is strictly contained in $\mathcal F[\mR]$.
\end{Lemma}
\begin{Proof}
  Considering  \lemmaref{lem:winmoveIsNTwo}, we assume
  towards a contradiction that there is a coordination-free 
  $\mP$-transducer $\mathcal T$ computing the won positions of the win-move game.
  Let $I = \set{ \boxupOne{move(a,b)} }$. Since
  $\mathcal T$ is coordination-free it computes the correct result 
  \lstinline{win(a)}
  on all
  networks $N$ and a partitioning policy $\PartPol$ 
  with only performing $k$ heartbeat transitions for some $k \in \mathbb N$.
  Choose the network
  that contains only a single node $n_0$. We observe that this implies $\PartPol
  \equiv \set{n_0}$.
  Now, consider input $I' = \set{ \boxupOne{move(a,b)}, \boxupOne{move(b,c)}}$ on the
  network $N' = \set{n_0,n_1}$. We choose a partitioning policy $\PartPol'$ that
  assigns
all ground atoms to $n_1$ except \lstinline{move(a,b)},
which is assigned to $n_0$.
  Consider stepping $\mathcal T$ on $n_0$ with $k$ heartbeat transitions.
  $\mathcal T$ will output \lstinline{win(a)} since its input and system
  relations \lstinline{move} and \lstinline{Local_MOVE} have the same contents
  as above (remember, $\mathcal T$ is deterministic if run only with heartbeat
  transitions). This is a contradiction since \lstinline{win(a)} is \emph{not} part
  of the correct result for $I'$.
\end{Proof}

We finish the proof of Theorem~\ref{thm:main} by showing that the last
containment in the hierarchy is strict.

\begin{Lemma}
Not all computable queries are coordination-free in model \mR, i.e.,
$\mathcal{F}[\mR] \subsetneq \mathcal{F}[\mA] = \mathcal{C}$.
\end{Lemma}

\begin{Proof}
Assume towards a contradiction that the query that decides whether a unary
relation $R$ is empty is in $\mathcal F[\mR]$. Following the
argument from the proof for \lemmaref{lem:winmoveNotNOne}, consider $I = \emptyset$
on the network with one node $n_0$; and then $I' = \set{ \boxupOne{R(a)} }$ together
with $\PartPol' \equiv \set{n_1}$ on the network $N'=\set{n_0,n_1}$.
\end{Proof}

 We point out that although win-move \emph{is} coordination-free in \mR, this
 does not directly imply that all well-founded \datalogneg is in $\mathcal F[\mR]$, even
 though ``win-move'' is a normal-form for well-founded Datalog \cite{Flum97}.
 The normal-form result for win-move as
 described in \cite{Flum97} requires a recursion-free semi-positive pre-processing
 step. It is not clear how, in general, this can be done in a coordination-free
 manner.

\section{Related Work}\label{sec:relatedWork}

\TODO{Make Ameloot more prominent: many proofs are inspired / similar to the
techniques used in \cite{Ameloot11}.}

Our work is inspired by Hellerstein's quest for logic-based
foundations of parallel, distributed, data-intensive computing
\cite{Hellerstein10}, and the search for suitable models of
computation and parallel complexity.
Hellerstein's model incorporates non-determinism via the choice
construct with an explicit representation of time within Datalog; we
use instead a bag of pending updates.
The paper by Alvaro et al.~\cite{alvaroconsistency} also studies
methods to ensure eventual consistency of distributed computations. In
particular, the authors analyze Dedalus programs (a version
of distributed Datalog) to identify points of non-monotonicity that
require coordination.  

Much of our technical approach is inspired by the paper by Ameloot et
al.~\cite{Ameloot11}, which introduced the framework of networks of
relational transducers and was the first to venture a formal notion of
coordination complexity.  We have shown that slight variations on the
model have a dramatic impact on the ensuing notion of coordination
complexity.

A more recent paper by Ameloot et al.~\cite{Ameloot2012} investigates
syntactic restrictions under which eventual consistency of networks of
relational transducers can be decided.  Our paper complements that work
by introducing disorderly evaluation of semi-monotone \DLx\ programs
and a compilation procedure for such programs guaranteed to create
eventually consistent transducer networks.

Our disorderly semantics is closely related to the so-called
``production-rule'' semantics of \datalogneg
\cite{abiteboul1991Extension,Vianu1997}. 
Recent work by Abiteboul et al.~\cite{abiteboul2011rule} also
investigates confluence and distributed systems, however without the
focus on coordination freeness.

The paper by Koutris and Suciu~\cite{Koutris11} presents a ``massively
parallel'' computation model (MP) for conjunctive queries that takes
data skew into account.  They show that in the MP model, the so-called
``tall-fat queries'' are precisely the ones that can be executed in a
single stage.  It would be interesting to investigate formal notions
of coordination that take into account data skew.

Finally, the paper by Brass et al.~\cite{BrassDFZ01} describes
transformation based strategies for incremental, bottom-up evaluation
of \datalogneg programs under the well-founded semantics
\cite{VanGelder93}. Some of their conditional rewriting techniques
resemble ours, but the paper does not address distributed or
coordination-free computations.

\section{Conclusion}
\label{sec:conclusion}

We have presented a syntactic fragment of \DLx, semi-monotone Datalog,
that lends itself to a disorderly, eventually consistent evaluation
strategy, while permitting certain kinds of negation.  We have shown
that the win-move game, the canonical example of a non-stratifiable
\datalogneg program, can be solved using a semi-monotone Datalog program,
and that the program can also be compiled into a relational transducer
network for distributed evaluation without resorting to ``global
synchronization barriers.'' 

We have also introduced several natural variations on relational
transducer networks~\cite{Ameloot11}, in which various degrees of
partitioning policy information are made available to participating
network nodes, and we used these models to study the notion of
coordination-freeness.  We showed that the classes of queries which
are coordination-free under these various models form a strict
hierarchy, highlighting the sensitivity of the formalization of
coordination-freeness by Ameloot et al.~to the precise details of how
partitioning policy information is made available to nodes.  The
win-move game, and our semi-monotone Datalog program for solving it, play
a starring role in separating two of these classes.

Like Ameloot et al.~\cite{Ameloot11}, we do not claim that the notions
of coordination-freeness developed here are definitive\footnote{Current events in American campaign finance law~\cite{Overby12}
suggest that confusion over the precise meaning of ``coordination'' is
not limited to the database theory community!}.  
However, our results can be interpreted as informing the ongoing development
of more robust notions of coordination-freeness.

\bibliographystyle{abbrv}

\appendix

\section{Proofs}

\subsection{Proofs for \secref{sec:disorderly}}

\begin{proof} (Theorem~\ref{thm:disorderlyTermination})
  We show the undecidability of all three problems via reduction from
  the problem of checking containment of (positive) Datalog programs,
  which is known to be undecidable~\cite{Shmueli93}, to deciding
  termination of \DLnegneg.  Let $P_1$ and $P_2$ be two Datalog
  programs with \edb{P_1} = \edb{P_2}, assume w.l.o.g.~that \idb{P_1}
  and \idb{P_2} are disjoint, and let $p_1$ (resp.~$p_2$) denote the
  distinguished output predicate of $P_1$ (resp.~$P_2$).

  \textit{Termination and eventual consistency}. Define $P$ to be the
  \DLx\ program that is the union of the rules of $P_1$ and $P_2$,
  along with
\[
\fbox{
  \boxup
  {toggle :- p_1(barX), !p_2(barX).}
  {!toggle :- toggle.}
}
\]
and with $\eidb{P} = \emptyset$.  It is easy to see that $P$ is
terminating under the disorderly semantics iff $P_1$ is contained in
$P_2$.  Since $P$ is also functional---the result being empty for any
input database instance---it follows that $P$ is also eventually
consistent iff $P_1$ is contained in $P_2$.

\textit{Functionality}.  Introduce a disjoint copy $P_1'$ of $P_1$
(with distinguished output predicate $p_1'$), and define $P$ to be the
\DLx\ program that is the union of the rules of $P_1$, $P_1'$, and
$P_2$, along with
\[
\fbox{
  \boxup
  {q(barX) :- p_1(barX), !p_1'(barX).}
  {!q(barX) :- p_2(barX).}
}
\]
where again $\eidb{P} = \emptyset$.  If $P_1$ is contained in $P_2$,
then the result of $P$ is empty on any source database, hence $P$ is
functional.  On the other hand, if $P_1$ is not contained in $P_2$,
then the order of firing the rules of $P_1$ and $P_1'$ matters, and
$P$ is not functional.
\end{proof}

\begin{Proof} (Theorem~\ref{thm:functional})
  Termination immediately follows from 
  the fact that each ground atom from $\ground{P,\cI}$ can only be changed
  (inserted into $\cI$ or deleted from $\cI$) once. 
  A finite number of changes implies termination since any (fair) trace has to
  reach a fixpoint.

  It remains to establish functionality.
  First, it will be convenient to assume that use of the {\bf
    request} case of the immediate successor relation is restricted
  to disallow requesting ``useless'' updates as follows: an
  insertion of $A'$ is requested in a state $(I,U)$ only if $A' \not\in
  I$ and $A' \not\in U$; conversely, a deletion of $\neg A'$ is
  requested only if $A' \in I$ and $\neg A' \not\in U$.  Using the
  semi-monotonicity of $P$, one can check that this restriction does
  not affect the result sets for $P$ under the disorderly semantics.

  Now, fix a semi-monotone \DLx\ program $P$ and input instance $\cI$.
  Under the amended definition of immediate successor given above, we
  claim (1) that any sequence $(\cI_0, \cU_0), (\cI_1, \cU_1), \dots$
  of states such that for each $i$, $(\cI_{i+1}, \cU_{i+1})$ is an
  immediate successor of $(\cI_i, \cU_i)$ using some rule in $P$ must
  be finite.  Indeed, this is a straightforward consequence of our
  earlier observation that the \IDBP\ relations only grow during
  computation, while the \IDBM\ relations only shrink (and the
  finiteness of \cI).

  Next, observe (2) that if $\cJ$ is in the result of applying $P$ to
  $\cI$ under the disorderly semantics, then $(\cJ, \emptyset)$ is an
  eventual successor of $(\cI, \emptyset)$.  To see this, suppose that
  $(\cJ, \cU)$ is an eventual successor of $(\cI, \emptyset)$.
  Let $(\cJ', \emptyset)$ be the eventual successor of $(\cJ, \cU)$
  reached by repeated application of the pending updates.  By
  Definition~\ref{def:disorderly}, since $\cJ$ is in the result set,
  it must be the case that $\cJ' = \cJ$.  Since
  $(\cJ', \emptyset)$ is also an eventual successor of $(\cI,
  \emptyset)$, we conclude that $(\cJ, \emptyset)$ is an eventual
  successor of $(\cI, \emptyset)$ as required.

  Then, we show (3) the following weak confluence property: any
  distinct immediate successors $(\cI_1, \cU_1)$ and $(\cI_2, \cU_2)$
  of a state $(\cI, \cU)$ have a common immediate successor $(\cJ,
  \cV)$.  Intuitively, whatever the change from $(\cI, \cU)$ to
  $(\cI_1, \cU_1)$, we can show that it still applies to $(\cI_2,
  \cU_2)$, and commutes with the change from $(\cI, \cU)$ to $(\cI_2,
  \cU_2)$.  We show this by an analysis of the nine possible cases.
  First, suppose that $(\cI_1, \cU_1)$ and $(\cI_2, \cU_2)$ follow as
  immediate successors of $(\cI, \cU)$ using just the {\bf insert}
  or {\bf delete} changes.  Then it is clear from inspecting the
  definition of immediate successor that the changes commute.  Next we
  consider the five remaining cases involving some use of {\bf
    request}.  Suppose that $(\cI_1, \cU_1)$ immediately succeeds
  $(\cI, \cU)$ using {\bf request}, with a rule $H \gets \forall
  \Xbar B_1, \dots, B_n$ and valuation $\nu$.  It is easy to see that
  if $(\cI_2, \cU_2)$ also succeeds $(\cI, \cU)$ using {\bf
    request}, then the changes commute; and that the same is true
  for {\bf insert} or {\bf delete}, using the semi-monotonicity
  of $P$.  The remaining cases are symmetric.

  Finally, we observe that (1) and (3) imply that the immediate
  consequence relation is strongly confluent.  Together with (2), this
  establishes the theorem.
\end{Proof}

\subsection{Proofs for \secref{rundisorderlyontransducers}}
\label{appendixproofsimulation}

\newcommand{\RepeatThingy}[3]{
\noindent \textbf{#1 \ref{#2}}\,\,\emph{#3}
}

\RepeatThingy{Lemma}{thmwoforall}{
  Let $Q$ be a query computed under the disorderly semantics by a
  semi-monotone, \fa-free \DLx program $P$ in which
  each rule has at most one negated (eidb)$^{*}$ atom.  Then there exists a
  type $\mP$ $\UCQneg$-transducer network which computes $Q$.
}
\\
\noindent $^\ast$ The proof for \lemmaref{thmwoforall} 
below assumes a stronger syntactic restriction
requiring each rule to contain only a single negative atom. The follow lemma shows
that this can safely be assumed: 
\begin{Lemma} Let $Q$ be a query computed under the disorderly semantics by a
  semi-monotone, \fa-free Datalog$^{\neg\neg}$ program $P$ in which
  each rule has at most one negated eidb atom. Then $Q$ can 
  be computed under the disorderly semantics by a
  semi-monotone, \fa-free Datalog$^{\neg\neg}$ program $P'$ in which
  each rule has at most one negated atom.
\end{Lemma}
\begin{Sketch}
  Let $r = (H \leftarrow body, \neg A) \in P$ be the rule with more than one negated
  atoms. Replace $r$ by $(H' \leftarrow body)$ and $(H \leftarrow H',\neg
  A)$ where $H' = R(\bar x)$ for $R$ being a fresh relation symbol and $\bar x$
  being all variables occuring in $body$. The resulting program is still
  semi-monotone and it is then easy to see that it still computes $Q$.
\end{Sketch}

\RepeatThingy{Lemma}{thmwforall}{
  Let $Q$ be a query computed by a semi-mon\-o\-tone \DLx program $P$
  under the disorderly semantics.  If all rules in $P$ are friendly, then there
  exists a type $\mR$ FO-transducer network which computes $Q$.
}
\\

All proofs are constructive. Starting from a semi-monotone program for
$Q$, we construct a consistent FO-transducer
$\mathcal T$ that computes $Q$ in model \mP.  We will see that the policy restriction to be
\emph{element-determined} (restricting $\mathcal T$ to be a type $\mR$
transducer) is important to allow multiple negated body atoms and
\fa-quantification while still guaranteeing that $\mathcal T$ computes
$Q$, \ie is deterministic under all allowed circumstances. The
differences for the proofs of \lemmaref{thmwoforall} and
\lemmaref{thmwforall} are only required in the proofs of
\Lemmaref{lemmalocalevaluable} and \Lemmaref{lemmafairrun}.

Given $Q$; let $P$ be the semi-monotone program computing $Q$.  As a
first step modify $P$ to $P'$ by adding one idb relation $R'$ for each
$R \in edb(P)$ via the rule:
\[
   \boxupOne{R'(barX) :- R(barX).}
\]
Also, in all other rules of $P$, replace \lstinline{R(barX)} by
\lstinline{R'(barX)}. Note that it is easy to see that $P'$ computes
the same query as $P$; also $P'$ is semi-monotone and satisfies the
requirements from \lemmaref{thmwoforall} (resp. \lemmaref{thmwforall}) if
$P$ did.

We then enrich $P'$ with the system-relations \Local for each $R\in
eidb(P)$ to obtain $P''$ in the following way: Consider each rule $r$
of $P'$ in turn. For each body atom $R(\bar x)$, add an additional
atom \Local$(\bar x)$ to the body of $r$ if $R$ is an eidb relation of
$P'$; if \Local$(\bar x)$ contains a \fa-quantified variable $x$,
replace $x$ with the variable $z$ as in the requirement of
\lemmaref{thmwforall}.

\begin{Example}
  Applied to the semi-monotone win-move program $\wmdmon$, this
  two-step transformation yields:
\[\;\;
\fbox{
\boxuppppppSix
  {*(0) space move'(X,Y) :- move(X,Y).}
  {*(1) space won(X) :- move'(X,Y), }
  {-- -- -- -- -- !may_win(Y), Local_MAYWIN(Y).}
  {*(2)!good_move(X,Y) :- won(Y), move'(X,Y).}
  {*(3)!may_win(X) :- forall Y !good_move(X,Y),}
  {-- -- -- -- -- Local_GOODMOVE(X,X), move'(X,_).}
}\;  (\wmdmonL)
\]
\end{Example}

Before describing the operations in $\mathcal T$, we develop some
programming techniques for transducers.  We first need to emulate
\emph{update-able eidb relations} in \mP-FO-transducers, since the
relations in $\Sin$ are read-only for the transducers. To do so, we
essentially program the transducer to---as a first action---copy the
data of eidb relations into the memory. During normal operation, we
refer to the memory version of the eidb relations. The construction is
a little tedious because we cannot control whether the transducer
performs heartbeat or network transitions in the beginning.  However,
with a number of boolean flags (\eg to indicate when the copy had
happened), and buffers for to-early-read network tuples, this can be
done. To simplify notation later on, we also copy the edb relations
into the memory of the transducers; these will not be changed during
the local computation.

We then need another programming technique: \emph{broadcast}. We
\emph{emulate} broadcast by always including the received fact in a
network transition in the facts to be sent out to the neighbors. Since
the network is connected every fact that is output on any transducer
will eventually be read by any other transducer in the network.

We now describe the operations of the \mP-FO-transducer $\mathcal
T$. The FO queries that implement that behavior are easy to envision.
$\mathcal T$ has the idb, eidb and edb relations of $P'$ in his
memory; the \Local relations are system relations. Note that edb and
eidb relations are initially partitioned according to
$H_{\PartPol,I}$.  All updates to idb and eidb relations will be
broadcasted during the run of the transducer network. During each
transition, $\mathcal T$ evaluates the FO-queries denoted by the
\emph{bodies} of the semi-monotone program $P''$ over its memory to
determine a set of updates.  These updates are (1) locally applied
(via the appropriate $Q^R_\tnins$ and $Q^R_ \tndel$ relations); (2)
also broadcasted (via appropriate $Q^R_\tnsnd$); and (3) output (via
$Q_\tnout$) if they are insertions for the designated idb output
predicate.
Since the relation-name $R$ already determines whether an update is an
insertion or a deletion the update ground atom $R(\bar x)$ can be
sent. On the receiver side, it is then put into appropriate
$Q^R_\tnins$ and $Q^R_\tndel$ relation based on whether $R$ is an idb
or eidb.

Fix a network $N$ and an arbitrary partitioning policy $\PartPol$.
Next, we will show a correspondence between (1) any run
$run(I,\mathcal T,N,H)$ of the constructed transducer $\mathcal T$ in
network $N$ on an arbitrary input $I$ over the schema $edb(P) \cup
eidb(P)$ with $H$ determined by $\PartPol$ and (2) a sequence of
eventual successors under the disorderly semantics of running $P'$ on
$I$ (note that the disorderly program we compare to is the \Local-free
version $P'$).

The main difference between the transducer network and the trace of
the disorderly program is that in the transducer network the program
$P''$ is run with \emph{local knowledge} whereas in the disorderly
semantics, an applicable body atom is evaluated based on the
``global'' database state. The next lemma states that, \emph{at
  least}, the transducer network is not performing any \emph{wrong}
updates, provided the syntactic restrictions from \lemmaref{thmwoforall}
and \lemmaref{thmwforall} are met for the respective transducer types.
We will use this lemma to subsequently show that the transducer is
deterministic; which will allow us to choose any order of
transitions. Which is then sufficient to show that the network
transducer is not \emph{missing} any derivations, \ie each fair run of
the transducer corresponds to a fair trace of the disorderly
semantics.

We first relate the global state \mState of a transducer network to a
single, database instance $M(\mState)$ over $\Smem$. For any $R \in
edb(P'') \cup idb(P'')$, $M(\mState)$ contains the \emph{union} of the
memory relations $R$ across the network.  For $R \in eidb(P'')$ we
take the \emph{intersection modulo scope}, that is a fact $R(\bar x)$
is in $M(\mState)$ iff it is present on \emph{all} nodes $v$ with $v
\in \PartPol(\,R(\bar x)\,)$, \ie that have $R(\bar x)$ in scope. We
ommit the obvious (but somewhat tedious) formal definition.
Furthermore, let $N(\mState)$ be the \emph{bag} containing the
bag-union of all local network buffers. %
\begin{Lemma} \label{lemmalocalevaluable} 
  Let \mState be an arbitrary global state of the transducer $\mathcal T$.
  Let $S_D :=
  (M(\mState),N(\mState))$ be a state for the disorderly semantics as
  defined in \secref{subsec:disorderlySemantics}.  Let $v$ be an
  arbitrary node in $N$ with $\mState(v) = (J,B)$, \ie the contents of
  its memory relations being $J$. Then, $R(\bar x) \in Q^R_\tnsnd(J)$
  implies that an insert/delete-update $\nu(H) = R(\bar x) / \neg
  R(\bar x)$ can be derived in $S_D$ according to $P'$.
\end{Lemma}
\begin{Proof} 
  Let $R(\bar x)$ be a ground atom with $R(\bar x) \in Q^R_\tnsnd(J)$;
  Since $Q^R_\tnsnd$ is the FO-query obtained from the bodies of
  $P''$, the needs to be a valuation $\nu$ such that $J \models
  \nu(body'')$ for a rule
  \beforeeqsqueeze
  \[
  (\neg)R(\bar x) \leftarrow body'' \quad \in P''.
  \]
  We need to show that not only $J \models \nu(body'')$ but also $M(\mState) \models \nu(body')$ with $body'$
  being the body of the corresponding rule in $P'$. Individually,
  consider all body atoms $A \in body'$. If $A$ is a positive atom,
  then clearly $M(\mState) \models \nu(A)$.  Now consider a negative
  body atom $\neg R(\bar x)$ not containing a \fa-quantified
  variable. According to our transformation from $P'$ to $P''$, also
  the atom \Local$(\bar x)$ is in $body''$; thus, $v$ has $R(\bar x)$
  in scope but does not contain $R(\bar x)$, which implies $R(\bar x)
  \not\in M(\mState)$. Without \fa-quantification, this completes the
  proof.  Note, that for this lemma the restriction to one negative
  body atom in \lemmaref{thmwoforall} is not required.  The existence of
  \fa-quantified body atoms requires the restrictions given in
  \lemmaref{thmwforall} and a restriction to type \mR
  transducers. W.l.o.g., consider a the single ``ground'' body atom
  with only one \fa-quantified variable. (It is easy to see that the argument
  holds for the other cases too).
  \lstinline{forall Z !R(c_1..c_k,Z,c___k+1__..c_n)} with
  \lstinline{c_i} being constants from the domain. Since $body''$ is
  \emph{friendly}, the exists at least one \lstinline{c_i}
  such that the non-ground atom had a variable $x \neq $\lstinline{Z}
  with $\nu(x) =$ \lstinline{c_i}.  Since $body''$ is guarded by
  \Local, $J\models$ \lstinline{Local_R(c_1..c_k,c_i,c___k+1__..c_n)},
  thus for the node $v$ under consideration $v \in
  \PartPol($\lstinline{R(c_1..c_k,c_i,c___k+1__..c_n)}). But since
  $\PartPol$ is \emph{element-determined}, we can conclude that for
  all $z$ ranging over the universe, $v
  \in \PartPol($\lstinline{R(c_1..c_k,}$z$\lstinline{,c___k+1__..c_n)}). Thus,
  $v$ is responsible for all relevant tuples which implies the goal
  $M(\mState) \models $
  \lstinline{forall Z !R(c_1..c_k,Z,c___k+1__..c_n)}.
\end{Proof}

We now show that any transition taken by the relational transducer can
be emulated by the disorderly semantics, in the following sense:
\begin{Lemma} \label{Lemmaeventualsuccessor} 
  If the transducer network $\mathcal T$ can transition from a state
  $\mState$ to $\mState'$, then $(M(\mState'),N(\mState'))$ is an
  eventual successor of $(M(\mState'),N(\mState'))$ under the
  disorderly semantics for $P'$.
\end{Lemma}
\begin{Sketch} 
  The proof is technical, but straightforward using
  \lemmaref{lemmalocalevaluable}.  Using multiple request and
  update rule invocations, the disorderly semantics can emulate the
  set-based FO-queries of a single transducer.  A key ingredient is
  the fact that if a node $v$ in the transducer network is the first
  (amongst all nodes) that derives a new idb fact (or delete an eidb
  fact), then $M(\mState)$ changes; if it was not the first then
  $M(\mState)$ does not change even though $\mState$ does. In this
  case, the change in pending network messages can be accomodated by
  the disorderly semantics since $P'$ is semi-monotone.
\end{Sketch}
It is important to \emph{also} show that the trace computed by the transducer network 
corresponds to a \emph{fair trace} of the disorderly semantics according to
\defref{fairtrace}.
\begin{Lemma} \label{lemmafairrun} Every fair run of $\mathcal T$ corresponds to a fair trace of $P'$
under the disorderly semantics.
\end{Lemma}
\begin{Sketch} Being a fair run ensures that every sent update is
  eventually applied. The more interesting part of the proof is to
  show condition (1) from \defref{fairtrace}. This is somewhat dual to
  \lemmaref{lemmalocalevaluable}.  We need to show that being able to
  derive an update in the global state implies that there is (\emph{or
    will be in a later state}) a node that can derive the same update
  based on its local knowledge.  That is, we need to show if
  $M(\mState) \models \nu(body')$ then there exists an eventual
  successor state $\mState'$ of $\mState$ that will be reached by any
  fair run of $\mathcal T$, for which there is a node $v \in N$ with
  memory contents $J$ such that $J \models \nu(body'')$.
  Fix a $\nu(body')$ with $M(\mState) \models \nu(body')$. If $body'$
  does not contain any negated atoms than $body'' = body'$. But now,
  the claim is easy to show since either $body'$ is a single edb atom
  (in which case we are done according to the definition of
  $M(\mState)$) or $body'$ is a join of \emph{only} idb relations; the
  claim then follows from the fact that we broadcast all idb-updates
  and the fact that $P'$ is semi-monotone.

  The more interesting cases are when $body'$ contains negated
  subgoals.  According to the same argument as above, we can w.l.o.g.,
  assume that on all nodes $v$ the positive body atoms are satisfied.
  Next, a differentiated analysis of (1) \lemmaref{thmwoforall} and (2)
  \lemmaref{thmwforall} is necessary. \textbf{(1)} $body'$ contains only
  one negated atom $\neg R(\bar x)$.  Since $M(\mState) \models \neg
  R(\bar x)$, none of the nodes $v \in \PartPol(R(\bar x))$ contain
  $R(\bar x)$ in their memory; consider one of these nodes $v$: both
  \Local$(\bar x)$ and $\neg R(\bar x)$ are true on $v$.  \textbf{(2)}
  Following the argument above, we see
  that for each $\neg R(\bar x),$\Local$\!\!(\bar x)$ pair in $body''$
  there exists a node $v \in N$ for which $\nu(body'')$ makes both
  true. We would like one single node $v'$ on which all of the pairs
  are true. Now, because the rule is friendly, all
  \lstinline{Local_R(...)} have one variable $x$ in common.  Since
  \lemmaref{thmwforall} requires a type \mR transducer and thus
  $\PartPol$ is element-determined, we can conclude that $v' =
  F(\nu(x))$ is the desired node.
\end{Sketch}
Since an initialization of the transducer network $\mathcal T$
corresponds to the initial disorderly state,
Lemmata~\ref{Lemmaeventualsuccessor}~and~\ref{lemmafairrun} prove
Theorem~\ref{thmwoforall} and \ref{thmwforall}.
\\

\RepeatThingy{Lemma}{thmcomposition}{
  Let $P$ be a projection program, and let $Q$ be a query as defined
  in Lemma~\ref{thmwforall}.  Then there exists a type $\mR$
  FO-transducer network which computes $Q \circ P$. %
}
\\

Consider the
semi-monotone program $P_Q$ for $Q$. Replace occurrences of $R \in
edb(P_Q)$ by their definition in $P$; the resulting program $P_Q'$ is
still semi-monotone. So, w.l.o.g., assume $P$ only defines projections for
relations $R$ that are used as eidb in $P_Q$. Also, for later, let $T$ be part
of $edb(P)$. To simplify the exposition,
consider only one of such $R$. We now modify the transducer
constructed in the proof for \lemmaref{thmwforall}. First, we add $R$ to the
eidb relations of the transducer (changing \Smem\ and \Smsg). We now change the 
procedure that initializes the memory relations from the input relation and
compute the projection query during the copy process.
While doing so, we also take the partioning policy $\PartPol$ into
consideration. Here, we use \lstinline{Local}$_T$ to get access to the function
$H$ as defined just above \defref{defmptransducer}.
In particular, for a projection rule \lstinline{R(barZ) :- S(barX)}, we 
use a UCQ as follows to decide what fragment of \lstinline{S} to copy into \lstinline{R}.
$$
\boxuppppFour
    {R(Z_1,..,Z_n) = S(barX), Local__T(Z_1,Z_1,..,Z_1).}
    {R(Z_1,..,Z_n) = S(barX), Local__T(Z_2,Z_2,..,Z_2).}
    {...}
    {R(Z_1,..,Z_n) = S(barX), Local__T(Z_n,Z_n,..,Z_n).}
$$
The second modification in the process of creating $\mathcal T$ is a change 
to the program $P_Q''$ (as by the proof
for \lemmaref{thmwforall}). We emulate the (non-existing system) relation 
\lstinline{Local_R(barZ)}
with \lstinline{Local}$_T$ as done above. 
It is easy to see that the emulated \lstinline{Local_R} and the contents of
\lstinline{R} corresponds to an allowed distribution policy for a type \mR
transducer. 

\end{document}